\documentclass[apj]{emulateapj}

\def\lae{\mathrel{<\kern-1.0em\lower0.9ex\hbox{$\sim$}}}
\def\gae{\mathrel{>\kern-1.0em\lower0.9ex\hbox{$\sim$}}}

\newcommand{\be}{\begin{equation}}
\newcommand{\ee}{\end{equation}}

\slugcomment{Accepted for publication in ApJ}
\shortauthors{JORD\'AN ET AL.}
\shorttitle{GLOBULAR CLUSTER HALF-LIGHT RADII}

\begin{document}

\title{The ACS Virgo Cluster Survey X. Half-light Radii of Globular
Clusters in Early-Type Galaxies: Environmental Dependencies and a Standard
Ruler for Distance Estimation\altaffilmark{1}}

\author{
Andr\'es Jord\'an\altaffilmark{2,3},
Patrick C\^ot\'e\altaffilmark{4,5},
John P. Blakeslee\altaffilmark{6},
Laura Ferrarese\altaffilmark{4,5},
Dean E. McLaughlin\altaffilmark{7},
Simona Mei\altaffilmark{6},
Eric W. Peng\altaffilmark{4},
John L. Tonry\altaffilmark{8},
David Merritt\altaffilmark{9},
Milo\v{s} Milosavljevi\'c\altaffilmark{10},
Craig L. Sarazin\altaffilmark{11}, 
Gregory R. Sivakoff\altaffilmark{11} and
Michael J. West\altaffilmark{12}
}

\begin{abstract}
We have measured half-light radii, $r_h$, for thousands of globular clusters (GCs)
belonging to the one hundred early-type galaxies 
observed in the {\it ACS Virgo Cluster Survey} and the elliptical galaxy NGC~4697.
An analysis of the dependencies
of the measured half-light radii on both the properties of the GCs themselves
and their host galaxies reveals that, in analogy with GCs in the Galaxy but
in a milder fashion,  
the average half-light radius increases 
with increasing galactocentric distance or, alternatively,
with decreasing galaxy surface brightness.
For the first time, we find that the average
half-light radius decreases with the host galaxy color. We
also show that there is no evidence for a variation of $r_h$ with 
the luminosity of the GCs. 
Finally, we find in agreement
with previous observations that the average $r_h$ 
depends on the color of GCs, with red GCs being $\sim 17 \%$ smaller
than their blue counterparts. We show that this difference
is probably a consequence of an intrinsic
mechanism, rather than projection effects, and that it is
in good agreement with the mechanism proposed in Jord\'an (2004).
We discuss these findings in light of two simple pictures
for the origin of the $r_h$ of GCs and show that both lead
to a behavior in rough agreement with the observations.
After accounting for the dependencies on galaxy color, galactocentric radius
and underlying surface brightness, 
we show that the average GC half-light radii $\langle r_h \rangle$ 
can be successfully used as a standard ruler for distance estimation.
We outline the methodology, provide a calibration for its use,
and discuss the prospects for this distance estimator with future observing facilities. 
We find 
$\langle r_h \rangle = 2.7\pm0.35$ pc for GCs with $(g-z)=1.2$ mag in a galaxy
with color $(g-z)_{\rm gal}=1.5$ mag and at an underlying surface $z$-band
brightness of $\mu_z = 21$ mag arcsec$^{-2}$. Using this technique,
we place an upper limit of 3.4~Mpc on the
1-$\sigma$ line-of-sight {\it depth} of the Virgo Cluster.
Finally, we examine the form of the $r_h$ distribution
for our sample galaxies and provide an analytic expression 
which successfully describes this distribution.
\end{abstract}


\altaffiltext{1}{Based on observations with the NASA/ESA
{\it Hubble Space Telescope}
obtained at the Space Telescope Science Institute, which is operated
by the Association
of Universities for Research in Astronomy, Inc., under
NASA contract NAS 5-26555}
\altaffiltext{2}{European Southern Observatory,
Karl-Schwarzschild-Str. 2, 85748 Garching bei M\"unchen, Germany; ajordan@eso.org}
\altaffiltext{3}{Astrophysics, Denys Wilkinson Building, University
of Oxford, 1 Keble Road, OX1 3RH, UK}
\altaffiltext{4}{Herzberg Institute of Astrophysics, 5071 W. Saanich Road, Victoria, 
BC V9E 2E7, Canada}
\altaffiltext{5}{Department of Physics \& Astronomy, 
Rutgers University, Piscataway, NJ 08854, USA}
\altaffiltext{6}{Department of Physics and Astronomy,
Johns Hopkins University, Baltimore, MD 21218, USA}
\altaffiltext{7}{Department of Physics \& Astronomy, University
of Leicester, Leicester, LE1 7RH, UK}
\altaffiltext{8}{Institute of Astronomy, University of Hawaii,
2680 Woodlawn Drive, Honolulu, HI 96822, USA}
\altaffiltext{9}{Department of Physics, Rochester Institute of Technology,
84 Lomb Memorial Drive, Rochester, NY 14623, USA}
\altaffiltext{10}{Theoretical Astrophysics, California Institute of Technology,
Pasadena, CA 91125, USA}
\altaffiltext{11}{Department of Astronomy, University of Virginia, P.O.
Box 3818, Charlottesville, VA 22903-0818, USA}
\altaffiltext{12}{Department of Physics \& Astronomy, University of Hawaii,
Hilo, HI 96720, USA}

\section{Introduction}
\label{sec:intro}

The half-light radius, $r_h$, represents one of the most fundamental
structural properties of globular clusters (GCs).
Numerical simulations have shown this quantity
to be fairly constant throughout the
dynamical evolution these stellar systems (Spitzer \& Thuan 1972; 
Lightman \& Shapiro 1978; Murphy, Cohn \& Hut 1990; 
Aarseth \& Heggie 1998). Moreover, there are reasons to believe that
the present day half-light radii are faithful records of the characteristic
sizes of the proto-GC clouds; thus, measurements of half-light radii
provide unique constraints on the properties of the GCs at the time
of their formation (Murray \& Lin 1992). 

Until recently, measurements of GC half-light radii were confined to
Local Group galaxies where the GCs could be resolved in ground-based
images. In the Milky Way, it has long been known that $r_h$ increases systematically
with galactocentric distance, $R_{gc}$ (van den Bergh 1956). van den
Bergh, Morbey \& Pazder (1991) found a scaling relation of the
form $r_h \propto R_{gc}^{0.5}$ based on an analysis of 98 Galactic 
GCs.  This situation changed dramatically with the launch of the {\it Hubble 
Space Telescope} ({\it HST}) which can partially resolve the spatial profiles of GCs well
beyond the Local Group; indeed, {\it HST} opened the study of 
GC sizes in external galaxies out to distances of $\sim$ 30 Mpc
(e.g., Kundu et~al.\ 1999; Puzia et~al.\ 1999;
Larsen et~al.\ 2001).
An interesting result to emerge from these studies is the
realization that the half-light radii
of metal-rich GCs in luminous early-type galaxies appear to be
$\sim 20\%$ smaller than those
of their metal-poor counterparts ---
a finding that has alternatively been
interpreted as the result of projection effects (Larsen \& Brodie 2003)
or the combined effects of mass-segregation and the dependence 
of stellar lifetimes on metallicity under the
assumption that their half-mass radius is constant
(Jord\'an 2004). Unfortunately, the fields
of view have been typically too small to explore possible dependencies
of $r_h$ on projected $R_{gc}$; only in one galaxy (NGC 4365) 
has a clear increase of
$r_h$ with projected $R_{gc}$ been found in the outermost regions
(Larsen et al. 2001).

The prospects for using the sizes of GCs for distance 
estimation dates back at least to the work of Shapley \& Sawyer (1927),
who state that
``measures of diameter [of GCs] are chiefly useful in the determination of
relative distances''. 
Recently, Kundu \& Whitmore (2001) used archival {\it HST}/WFPC2 observations
for a sample of 28 galaxies to argue that the median half-light radii
for GCs in early-type galaxies is constant at 
$\langle r_h \rangle =2.36\pm0.4$~pc.  This led them to suggest that
$\langle r_h \rangle $ could be used as a distance indicator, although
they noted that their sample showed a tendency for the physical size of
GCs to increase with galaxy distance, a trend they attributed to instrumental effects.

In this paper, we use the GCs from our large and homogeneous survey of early-type
galaxies in the Virgo Cluster to explore the dependencies of half-light radii
on various properties of the GCs and their host galaxies. We describe the observations
and data reductions in \S2. An exploration of the dependencies of half-light
radii on the properties of the GCs and  their host galaxies is given in \S3. 
Theoretical implications of the measured half-light radii are briefly discussed
in \S4. After accounting for various environmental dependencies,
we show in \S5 that $\langle r_h \rangle $ can be successfully used as a distance indicator
and we provide a method and calibration for its use. 
In \S6, we study the form of the observed 
$r_h$ distribution in early-type galaxies and
present an analytic expression
which successfully describes this distribution across our sample. We
summarize and conclude in \S7.
An appendix gives a description of the algorithm used to measure structural
parameters, including half-light radii, for GCs in the ACS Virgo Cluster Survey.

\section{Observations}
\label{sec:obs}

A sample of 100 early-type galaxies in the Virgo cluster 
was observed as part of the ACS
Virgo Cluster Survey (ACSVCS; C\^ot\'e et~al. 2004). Each galaxy
was observed in the F475W ($\simeq$ Sloan $g$) and F850LP
($\simeq$ Sloan $z$) bandpasses for a total of 
720~s and 1210~s respectively. Additionally, the
X-ray faint early-type galaxy NGC~4697 was observed with an identical 
strategy as part of a joint {\it Chandra}-{\it HST} program (GO-10003;
PI = C.L. Sarazin). Data
reductions were performed as described in Jord\'an et~al.\ (2004a). For all
GC candidates, a half-light radius is measured by fitting PSF-convolved
King (1966) models to the observed light distribution. The
methodology used to measure $r_h$ is described in Appendix~A.
Profile fitting 
is carried out independently in both the F475W and F850LP filters. For
this work, we estimate 
$r_h$ by averaging the results measured in the two bandpasses. Having two independent
measurements in different bands lets us estimate systematic uncertainties
arising from the adopted PSF models. We find that the measurements
in the two bands have systematic differences of  $\sim 0.05$ WFC pixels,
or $\sim 0.2$ pc at the distance of Virgo (Mei et~al. 2005b). We have thus
added in quadrature a systematic error of $0.05$ pixels, or $0\farcs 0025$, 
to the measured $r_h$ values.

The construction of a clean and complete GC catalog is an essential but difficult
part of all studies of extragalactic GC systems. While the full details of the GC
selection process will be described elsewhere, along with the
presentation of the GC catalogs (Jord\'an et~al., in preparation), 
we briefly outline the technique here. 
Making use of the measured $r_h$ and $z$-band magnitude and catalogs of
the expected contamination taken from blank control fields
and constructed for each galaxy individually, we
select GCs via maximum likelihood estimation of a 
mixture-model of GCs and contaminants. Due to the difficulty of
distinguishing extended GCs from background galaxies, sources
with $r_h > 0\farcs128$ ($\sim 10$~pc at the mean Virgo distance) are
eliminated as GC candidates before estimating the mixture model.
This method assigns to each candidate a probability, $p_{\rm gc}$, that it 
is a GC. In this work, we consider only those sources for which $p_{\rm gc} \geq 0.5$.

When analysing the $r_h$ measurements, it is crucial to minimize or avoid
any selection and contamination effects. In order to create a clean sample
for structural parameters studies, we have restricted our sample to
GC candidates to those with
\begin{equation}
\begin{array}{rrcrll}
& & z & \le & 22.9~{\rm mag} \\
%& & r_h & \le & 10~{\rm pc} \\
0.6 & \le & (g-z) & \le & 1.7~{\rm mag} \\
\end{array}
\label{sel1}
\end{equation}
Note that the selection on magnitude corresponds roughly
to the expected turnover
of the GC luminosity function at the distance of Virgo.
Additionally, we have imposed on {\it each program galaxy}
a selection on projected galactocentric distance.
An {\it inner cut}, determined with the help of our
artificial objects test, ensures that the underlying surface
brightness of the galaxy is such that our completeness for objects
with $r_h = 0\farcs128$ 
is greater than $90\%$.
An {\it outer cut}, determined
using our control fields, ensures that the expected 
contamination in our GC samples from background galaxies is at 
most $5\%$ of the objects in the sample. The average expected contamination
for the samples is $\sim 3\%$, with the values increasing with decreasing 
luminosity: i.e., from $\sim 1\%$ for the giants to $\sim 5\%$ for the least
luminous dwarfs. For our analysis, we considered only those galaxies whose
final GC samples contained five or more objects.\footnote{
Additionally, four faint galaxies located close to luminous giants were
eliminated since their own GC systems are overwhelmed by those of their
companions. These galaxies are VCC~1297, VCC~1199, VCC~1192 and VCC~1327.}
This left a final sample of 67 galaxies including NGC~4697.
The typical random uncertainties for the measured half-light radii are
$\approx 0\farcs003 \approx 0.24$~pc. 

As part of the ACS Virgo Cluster Survey, we have measured distances for
most of our target galaxies using the method of surface brightness
fluctuations (SBF; Tonry \& Schneider 1988). The reduction procedures for
SBF measurements, feasibility simulations for our observing configuration,
and the calibration of the $z$-band SBF method have been presented in 
Mei et~al.\ (2005ab). The complete catalog of SBF distances will be
presented in Mei et~al. (2005c). With distances to most our targets in hand, 
we may convert the observed half-light radii from angular into physical
units, and thereby examine directly the physical sizes of the GCs.
In what follows, we consider only those GCs which belong to
the galaxies with measured SBF distances. Our final sample is 
summarized in Table~\ref{tab:rhmean}, which lists the 67 galaxies upon
which our analysis is based, the blue magnitude, $B_g$, from the Virgo
Cluster Catalog of Binggeli, Sandage \& Tammann (1985), the
number of GCs satisfying the above selection criteria, 
the mean GC half-light radius in arcseconds and 
its uncertainty (not including systematic uncertainties)
and alternative names for our sample galaxies.

\begin{deluxetable}{crrrc}
\tablecaption{Mean Half-Light Radii for ACS Virgo Cluster Survey galaxies 
and NGC~4697\label{tab:rhmean}}
\tablewidth{0pt}
\tablehead{
\colhead{VCC} &
\colhead{$B_g$} & 
\colhead{$N_{\rm gc}$} &
\colhead{$\langle r_h \rangle$} &
\colhead{Other Names}
\\
\colhead{} &
\colhead{(mag)} &
\colhead{} &
\colhead{($\arcsec$)} &
\colhead{}
}
\startdata
1226 &   9.31 & 392 &$ 0.033\pm 0.001$ & N4472, M49 \\
1316 &   9.58 & 907 &$ 0.032\pm 0.000$ & N4486, M87 \\
1978 &   9.81 & 398 &$ 0.029\pm 0.000$ & N4649, M60 \\
881 &  10.06 & 199 &$ 0.036\pm 0.001$ & N4406, M86 \\
798 &  10.09 & 211 &$ 0.035\pm 0.001$ & N4382, M85 \\
763 &  10.26 & 257 &$ 0.030\pm 0.001$ & N4374, M84 \\
731 &  10.51 & 333 &$ 0.024\pm 0.000$ & N4365 \\
1903 &  10.76 & 190 &$ 0.033\pm 0.001$ & N4621, M59 \\
1632 &  10.78 & 244 &$ 0.031\pm 0.001$ & N4552, M89 \\
1231 &  11.10 & 136 &$ 0.034\pm 0.001$ & N4473 \\
1154 &  11.37 & 93 &$ 0.031\pm 0.001$ & N4459 \\
1062 &  11.40 & 98 &$ 0.038\pm 0.001$ & N4442 \\
2092 &  11.51 & 41 &$ 0.037\pm 0.002$ & N4754 \\
759 &  11.80 & 86 &$ 0.036\pm 0.001$ & N4371 \\
369 &  11.80 & 115 &$ 0.037\pm 0.001$ & N4267 \\
1692 &  11.82 & 56 &$ 0.038\pm 0.002$ & N4570 \\
2000 &  11.94 & 116 &$ 0.037\pm 0.001$ & N4660 \\
1664 &  12.02 & 79 &$ 0.036\pm 0.001$ & N4564 \\
944 &  12.08 & 44 &$ 0.033\pm 0.001$ & N4417 \\
1938 &  12.11 & 51 &$ 0.041\pm 0.002$ & N4638 \\
1279 &  12.15 & 74 &$ 0.039\pm 0.002$ & N4478 \\
355 &  12.41 & 8 &$ 0.036\pm 0.003$ & N4262 \\
1619 &  12.50 & 17 &$ 0.037\pm 0.002$ & N4550 \\
1883 &  12.57 & 32 &$ 0.039\pm 0.002$ & N4612 \\
1242 &  12.60 & 63 &$ 0.036\pm 0.001$ & N4474 \\
784 &  12.67 & 15 &$ 0.033\pm 0.001$ & N4379 \\
1537 &  12.70 & 8 &$ 0.032\pm 0.003$ & N4528 \\
778 &  12.72 & 25 &$ 0.030\pm 0.002$ & N4377 \\
828 &  12.84 & 16 &$ 0.032\pm 0.002$ & N4387 \\
1321 &  12.84 & 14 &$ 0.034\pm 0.002$ & N4489 \\
1630 &  12.91 & 16 &$ 0.028\pm 0.002$ & N4551 \\
1250 &  12.91 & 8 &$ 0.041\pm 0.007$ & N4476 \\
1146 &  12.93 & 36 &$ 0.038\pm 0.002$ & N4458 \\
1303 &  13.10 & 21 &$ 0.033\pm 0.001$ & N4483 \\
1913 &  13.22 & 31 &$ 0.043\pm 0.002$ & N4623 \\
1475 &  13.36 & 38 &$ 0.038\pm 0.003$ & N4515 \\
1178 &  13.37 & 51 &$ 0.040\pm 0.002$ & N4464 \\
1283 &  13.45 & 22 &$ 0.032\pm 0.002$ & N4479 \\
1261 &  13.56 & 17 &$ 0.034\pm 0.002$ & N4482 \\
698 &  13.60 & 56 &$ 0.039\pm 0.002$ & N4352 \\
1422 &  13.64 & 21 &$ 0.044\pm 0.003$ & I3468 \\
9 &  13.93 & 9 &$ 0.043\pm 0.003$ & I3019 \\
1910 &  14.17 & 17 &$ 0.039\pm 0.003$ & I809 \\
856 &  14.25 & 21 &$ 0.046\pm 0.003$ & I3328 \\
140 &  14.30 & 7 &$ 0.038\pm 0.001$ & I3065 \\
1087 &  14.31 & 29 &$ 0.042\pm 0.002$ & I3381 \\
1861 &  14.37 & 25 &$ 0.045\pm 0.002$ & I3652 \\
543 &  14.39 & 10 &$ 0.043\pm 0.003$ & U7436 \\
1431 &  14.51 & 28 &$ 0.036\pm 0.001$ & I3470 \\
1528 &  14.51 & 24 &$ 0.039\pm 0.002$ & I3501 \\
437 &  14.54 & 16 &$ 0.054\pm 0.006$ & U7399A \\
2019 &  14.55 & 10 &$ 0.047\pm 0.004$ & I3735 \\
1545 &  14.96 & 19 &$ 0.043\pm 0.003$ & I3509	 \\
230 &  15.20 & 11 &$ 0.039\pm 0.003$ & I3101	 \\
1828 &  15.33 & 7 &$ 0.060\pm 0.005$ & I3635	 \\
1407 &  15.49 & 11 &$ 0.044\pm 0.003$ & I3461	 \\
1539 &  15.68 & 17 &$ 0.043\pm 0.002$ &   \\
\hline
NGC4697 & 10.14  & 229  & $ 0.051\pm 0.001$ & \\
\enddata
\end{deluxetable}

\section{Environmental Dependencies}
\label{sec:depend}

As mentioned in \S1, GC half-light radii are believed to reflect
the physical conditions of the proto-GC clouds at the time of their 
formation. Thus, any systematic variations in half-light radius can provide
potentially powerful insights into the mechanism(s) of GC and
galaxy formation.

The various factors that can, in principle, affect the half-light radii
of GCs can be naturally
divided into factors that change $r_h$ individually
in each GC, factors that affect the GC system {\it within}
a galaxy, and factors that change it {\it across} galaxies. In what follows,
we shall refer to these factors as {\it internal},  {\it local} 
and {\it global} factors. The variables we will use to parameterize
internal effects are the GC colors, $(g-z)_{\rm gc}$, and $z$-band 
magnitudes. Local factors will be parameterized using the variables
$\log (r/r_e)$, where $r_e$ is the effective radius obtained from our
isophotal analysis (Ferrarese et~al.\ 2005, in preparation), and the 
underlying $z$-band surface brightness of the galaxy, $\mu_z$, at 
the position of each GC.
Note that $\mu_z$ is preferred over its $g$-band counterpart because
it is a better tracer of the underlying stellar mass and is
less sensitive to dust obscuration.
Finally, global parameters are taken to be the absolute blue
magnitude of the host galaxy, $M_B$, and its corresponding color, 
$(g-z)_{\rm gal}$ (Binggeli, Sandage \& Tammann 1985;
C\^ot\'e et~al. 2004; Ferrarese et~al. 2005).\footnote{For NGC~4697,
which is not a member of the Virgo Cluster and hence does not appear
in the catalog of Binggeli, Sandage \& Tammann (1985),
we use the integrated blue magnitude from NED: B = 10.14 mag. }
We use our SBF distance moduli (Mei et~al. 2005bc) to calculate 
absolute magnitudes on a galaxy-by-galaxy basis.

Many of these properties are highly correlated, 
so the extraction of the intrinsic dependence on each of them is
not entirely straightforward. For example, not only are
$M_B$ and $(g-z)_{\rm gal}$ correlated, but both quantities
are themselves correlated with the mean color of the GC systems 
and the fraction of GCs which belong to the metal-rich component
(with brighter and redder systems having a higher fraction
of metal-rich GCs; e.g., Peng et~al. 2005). 
Moreover, the dependence of the average half-light radius, 
$\langle r_h \rangle$, on $(g-z)_{\rm gc}$
(see below) will affect the inferred dependence of 
$\langle r_h \rangle$ on $\mu_z$ through differing 
proportions of metal-rich and metal-poor GCs as
a function of radius: i.e., red and blue GCs  
are often found to have different spatial distributions,
with the latter having a more extended spatial 
distribution (Kissler-Patig et~al.\ 1997; 
Lee, Kim \& Geisler 1998; C\^ot\'e et~al.\ 2001; 
Dirsch et~al.\ 2003).\footnote{
While we will refer throughout this work to $\langle r_h \rangle$ as
the average or median of $r_h$, we actually calculate 
its value using the biweight
location estimator (Beers et~al.\ 1990). }
Likewise, the dependence of $\langle r_h \rangle$ on
$(g-z)_{\rm gc}$ will also influence the 
dependence of $\langle r_h \rangle$ on $M_B$, as more
luminous galaxies have redder GC systems.

Our approach will therefore be to first isolate the various dependencies
(i.e., internal, local and global) on environment,
and then to ``correct" the observed $r_h$ so that
subsequent correlations are unaffected the preceeding
dependencies. We will indicate the successively corrected 
half-light radii by adding primes (i.e., $r'_h, r''_h$)
and will denote the final corrected $r_h$ (in which all 
environmental dependencies have been accounted for) 
with a hat: $\hat{r}_h$. Our goals in this portion of
the analysis are twofold. First, the individual correlations
hold potentially important clues to the processes of
GC formation and evolution (\S~\ref{sec:th}). And
second, we will show that --- once the various dependencies
on GC color, galaxy color and underlying surface brightness
are taken into account --- $\langle \hat{r}_h \rangle$ is 
remarkably constant across our sample, and thus appears to
be a promising ``standard ruler" for distance estimation
(\S~\ref{sec:distance}).

\subsection{Internal Factors}
\label{sec:dep:int}

\begin{figure}
\plotone{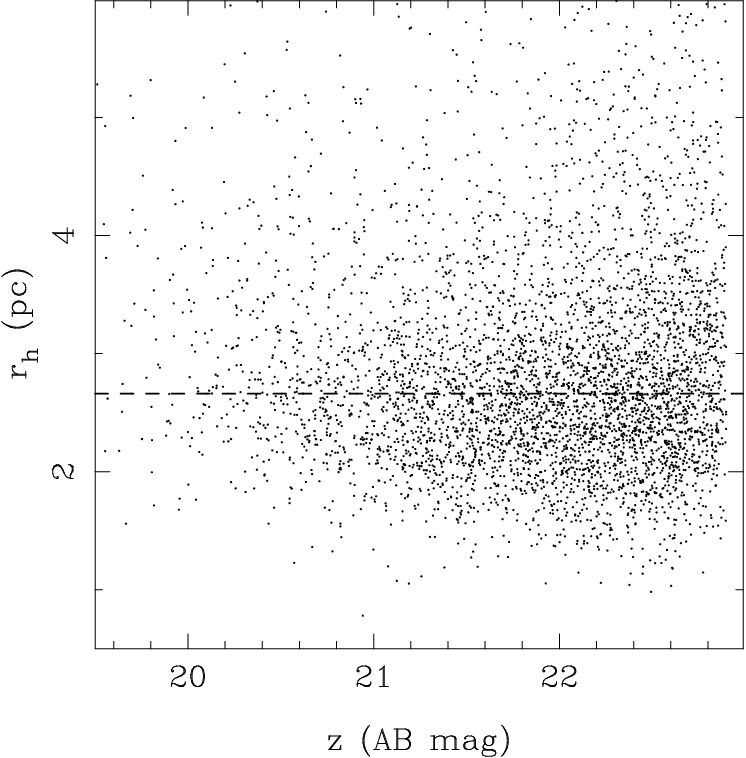}
\caption[]{Half-light radius, $r_h$, plotted as a function
of $z$-band magnitude for the combined GCs of
our sample galaxies. The data are consistent with $\langle r_h \rangle$ 
being independent of magnitude. The dashed line marks the median
value of $\langle r_h \rangle = 2.66$ pc. 
\label{fig:rhz}
}
\end{figure}

\begin{figure}
\plotone{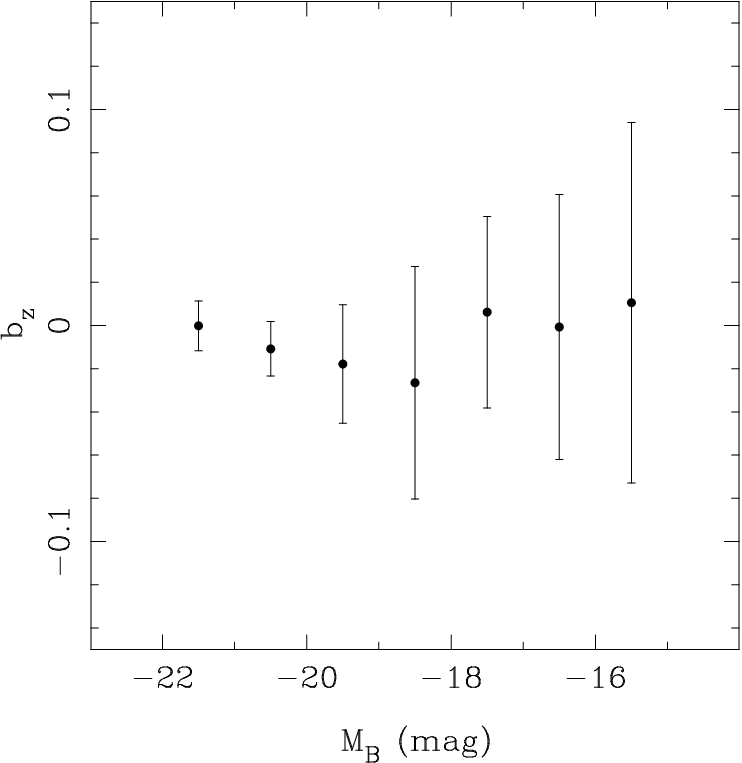}
\caption[]{Average coefficient $b_z$ calculated in one magnitude bins
for galaxies in the ACS Virgo Cluster Survey, plotted against their 
absolute blue magnitude, $M_B$. The $b_z$ have been obtained from a linear
fit of the relation $\log r_h = b_zz+a$ to the GC data for each galaxy. 
\label{fig:rhz_b}
}
\end{figure}

One of the most remarkable properties of GCs is that their
size appear to be independent of their mass: i.e., studies of the
Galactic GC system show that $r_h \propto M_{\rm gc}^0$ 
(McLaughlin 2000).
In fact, this relation is equivalent to one of the defining
equations of the ``fundamental plane'' of GCs, namely,
that GC binding energy scales with luminosity
as $E_{{\rm gc}} \propto L_{\rm gc}^{2.04}$ (McLaughlin 2000).
In Figure~\ref{fig:rhz},
we show $r_h$ versus $z$-band magnitude for our sample of GCs.
A regresssion analysis shows that the linear coefficient 
in the function $r_h = b_z z + a_z$
is not significant for this sample, meaning that, as expected,
GC radii in the ACSVCS are independent of luminosity. 
We have also performed fits of this same functional
form for all galaxies individually. The resulting average
linear terms as a function of absolute magnitude
are shown in Figure~\ref{fig:rhz_b}, which gives the weighted
average slope in each magnitude bin. It is apparent
that, across the factor of 350 in luminosity spanned by our
sample galaxies, there is no evidence for a variation in the
GC size with luminosity.
It should be noted that this statement applies only to those clusters
brighter than the turnover of the GC luminosity function ($z \le 22.9$)
due to the construction of the sample. To the extent that the $z$-band
luminosity reflects GC mass, this finding provides strong evidence 
that $r_h \propto M_{\rm gc}^0$.

On the other hand, previous {\it HST} studies of bright ellipticals have shown
that, in the mean, the half-light radii of GCs, $\langle r_h \rangle$,
depends on their color $(g-z)_{\rm gc}$, with the red GCs being 
$\approx 20\%$ smaller than their blue counterparts 
(Kundu \& Whitmore 1998; Kundu et~al. 1999; Puzia et~al. 
1999; Larsen, Forbes \& Brodie 2001; Barmby, Holland \& 
Huchra 2002; Larsen et~al. 2001).
As mentioned above, this color dependence will have implications
for other $r_h$ correlations: in particular,
it will affect correlations between $\langle r_h \rangle$
and $M_B$, $(g-z)_{\rm gal}$, $\log(r/r_e)$ and $\mu_z$. 
To avoid these complications, we choose to temporarily
restrict the GC sample to blue clusters only, with
$(g-z)_{\rm gc} < 1.05$. Transforming from color to metallicity
as in Jord\'an et~al. (2004b), this cut corresponds to a 
metallicity selection of [Fe/H] $\lae -1.25$.
Once we have characterized the remaining dependencies,
we will return to address the issue of the color dependence
of GC half-light radii. While
requiring the GCs to lie in a restricted range of 
color will not, in principle, entirely eliminate
any dependency, it will certainly reduce it to 
a very small level.
We note that the models presented in Jord\'an (2004)
for the variation of $r_h$ with $(g-z)_{\rm gc}$ --- which 
interpret the observed difference in size between red and 
blue GCs as a joint consequence of mass segregation and depedence of
stellar lifetimes on metallicity --- predict very little
variation for GCs with [Fe/H] $\lesssim -1$.

\subsection{Local Factors}
\label{subs:local}

\begin{figure}
\plotone{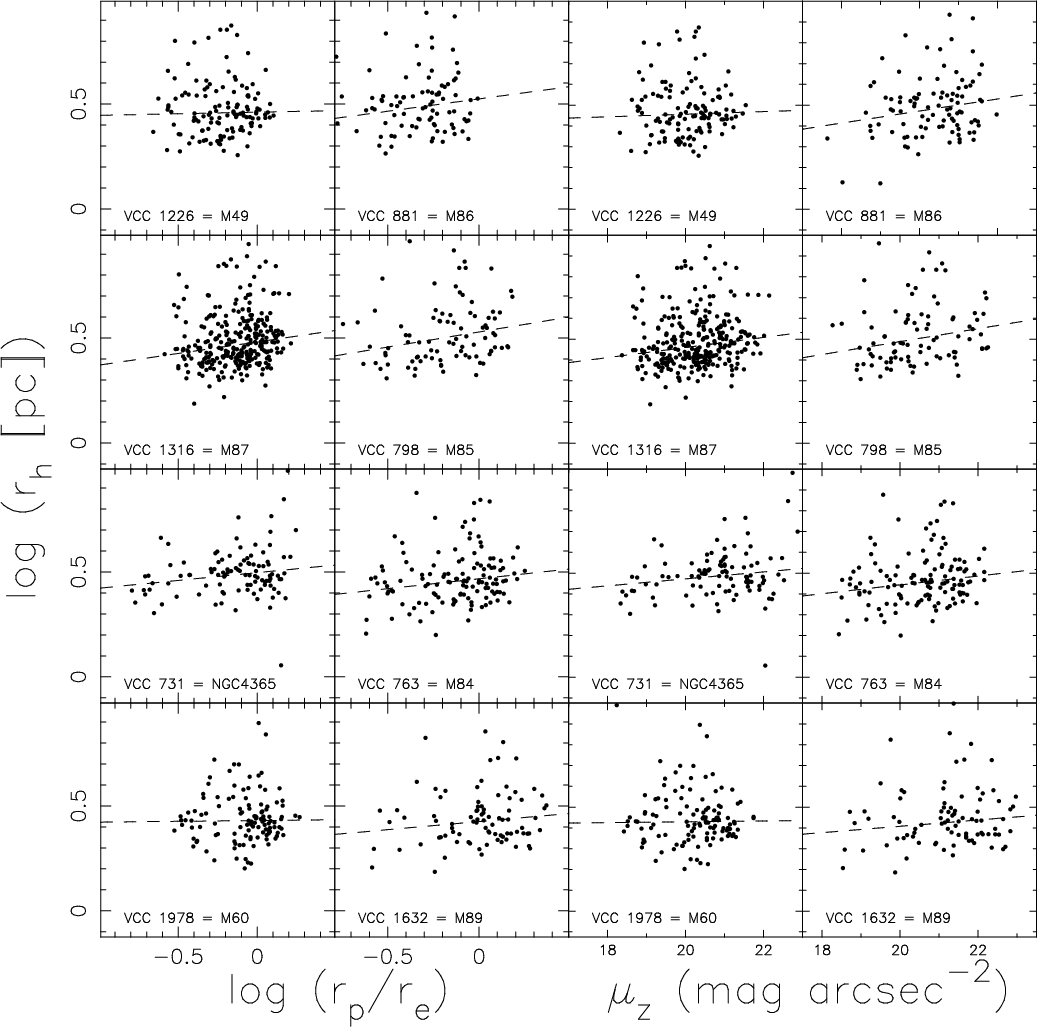}
\caption[]{({\it Two Leftmost Columns}) Logarithm of GC
half-light radius plotted as
a function of $\log (r_p/r_e)$, where $r_e$ is the galaxy
effective radius measured from the ACS/WFC images. Only GCs 
with $(g-z)_{\rm gc} < 1.05$ and belonging to the eight brightest
galaxies are shown in this figure. The dashed
lines indicate the best fit-linear relations. Galaxy
names are indicated in each panel.
({\it Two Rightmost Columns})
Logarithm of GC half-light radius plotted as
a function of the galaxy surface brightness in the $z$-band,
$\mu_z$. Only GCs
with $(g-z)_{\rm gc} < 1.05$ and beloginging to the eight brightest
galaxies are shown in this figure. The dashed
lines indicate the best fit-linear relations. Galaxy
names are indicated in each panel.
\label{fig:rh_rre_bright}
}
\end{figure}

\begin{figure}
\plottwo{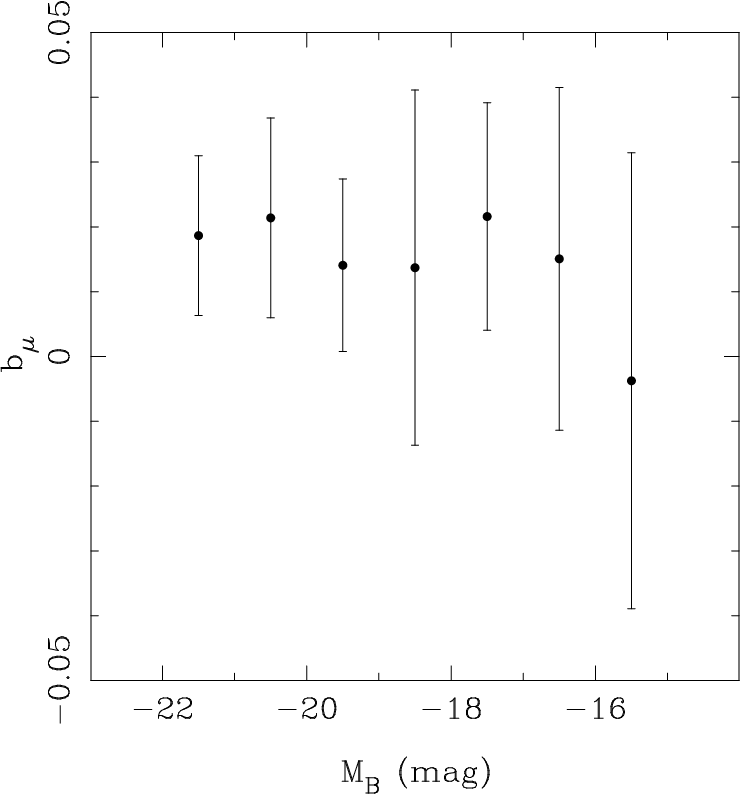}{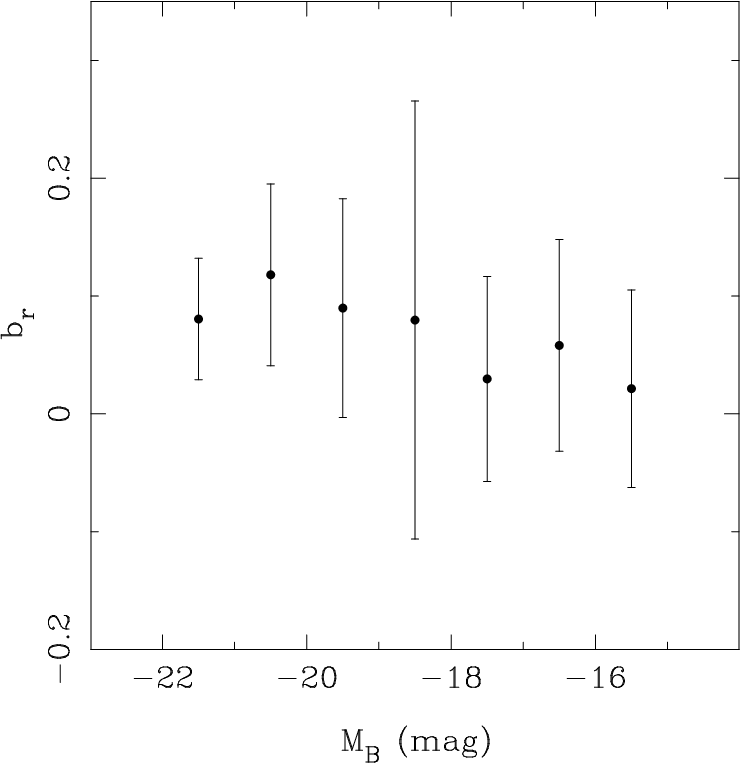}
\caption[]{ ({\it Left}) Average coefficient $b_\mu$ calculated in one magnitude bins
for galaxies in the ACS Virgo Cluster Survey, plotted against their
absolute blue magnitude, $M_B$. The $b_z$ have been obtained from a linear
fit of the relation $\log r_h = b_\mu \mu_z+a$ to the GC system of each galaxy.
({\it Right}) 
Average coefficient $b_r$ calculated in one magnitude bins
for galaxies in the ACS Virgo Cluster Survey, plotted against their
absolute blue magnitude, $M_B$. The $b_z$ have been obtained from a linear
fit of the relation $\log r_h = b_r \log(r_p/r_e) +a$ to the GC system of each galaxy.
\label{fig:bb}
}
\end{figure}

Using our restricted sample of blue GCs, we now search 
for correlations between the observed $r_h$ with projected galactocentric
radius, $r_p$, and local galaxy surface brightness, $\mu_z$.
In the two left columns of Figure~\ref{fig:rh_rre_bright}, 
we show $\log r_h$ versus
$r_p$ in units of the effective radius $r_e$ for the  eight
brightest galaxies in our sample; in the two right columns, we 
show $\log r_h$ versus $\mu_z$ for these same eight galaxies.
It is clear that the $r_h$ increases with both quantities, albeit
rather mildly.
We have performed linear fits of the form
$\log r_h = a_r + b_r\log (r/r_e)$ and
$\log r_h = a_\mu + b_\mu\mu_z$
for all galaxies in our sample. We show the resulting values
$b_r$ and $b_\mu$, plotted as a function
of galaxy absolute magnitude, $M_B$, in Figure~\ref{fig:bb}. We conclude
that most galaxies in the sample show evidence for an
increase in $r_h$ with increasing $r_p/r_e$ 
and decreasing surface brightness. A galaxy luminosity term
included in the regression analysis
is not significant at $99\%$ level in
either case, so we conclude that a single slope sufficiently describes 
these increases across the entire sample. The weighted average 
slopes for the fitted relations
are $b_\mu = 0.016\pm0.003$ and $b_r=0.07\pm0.01$, which we will take as the 
coefficients of a universal relation for
early-type galaxies.\footnote{If all GCs, rather than just the
blue ones, are used in the analysis, the coefficients change
only slightly, to $b_\mu = 0.017$ and $b_r=0.09$. The direction
of the change is as expected from the combination
of smaller $\langle r_h \rangle$ and less extended
spatial distribution for redder GCs.}

A slope of $b_r\approx 0.07$ is much lower than that found
for GCs in our Galaxy, $r_h \propto R_{\rm gc}^{0.5}$, where
$R_{\rm gc}$ is the Galactocentric distance. However,
these relations are not directly comparable because we observe
only the projected galactocentric radii, $r_p$, for the GCs in
our target galaxies. To compare with the Galactic relation, we
have performed 100 random projections of the Galactic GC system
using the positional data from Harris (1996), simulating what 
would have been observed with {\it HST}/ACS if the Milky Way 
was located at the distance of Virgo, and
the GC sample was selected in the same fashion as
for the ACSVCS targets. Only GCs with $M_V < -7.4$ mag 
(i.e. only the brighter half) were used in order
to reproduce the magnitude cut of our sample. 
To each simulated dataset
we have fit a linear relation as above to find
$b_r$, and recorded the result. These simulations shows
that our Galaxy would have been measured to 
have $b_{r,MW} \sim 0.3$, with a dispersion around this
value of $\sim 0.15$. Thus, the slope we find
for GCs in  early-types in Virgo is significantly
shallower than that for GCs in the Galaxy. 
While we will not explore this issue further in this work, 
it is an interesting discrepancy that may point to 
a difference in the formation or dynamical evolution of
GC systems in early- versus late-type galaxies, or perhaps,
in group versus cluster environments.

While either $\log (r_p/r_e)$ and $\mu_z$ may be use to parameterize the
outward decline in $r_h$, it is clear that they are not independent 
variables: i.e., quite generally, $\mu \propto \log F(r_p/r_e)$, where
$F$ is an arbitrary continuous 
function with finite total integral. This does not make
the descriptions in terms of the two variables necessarily
equivalent but it renders them correlated and forces the choice
of one. We choose to work with $\mu_z$ as the 
independent variable since it avoids additional uncertainties that 
arise due to errors in the determination of $r_e$,  
and in the assumption of spherical symmetry (which is a poor assumption
for a number of our program galaxies). Moreover, the use of $\mu_z$ 
accounts implicitly for the non-homologous nature of the galaxy profiles 
(e.g., Caon, Capaccioli \& D'Onofrio 1993), a behavior which a simple
power-law cannot capture. 

Using the relation between $r_h$ and $\mu_z$, we correct the value of
$r_h$ measured for each GC to the value expected at an underlying
galaxy surface brightness of $\mu_z=21$~mag~arcsec$^{-2}$. The
resulting corrected half-light radius $r'_h$ is given then by
\begin{equation}
r'_h \equiv r_h 10^{-0.016(\mu_z-21)}.
\label{eq:rprime}
\end{equation}

Note  that the form of the correction tacitly assumes that only the mean of the 
$r_h$ distribution is changing according to the derived relation.

\subsection{Global Factors}

\begin{figure}
\plotone{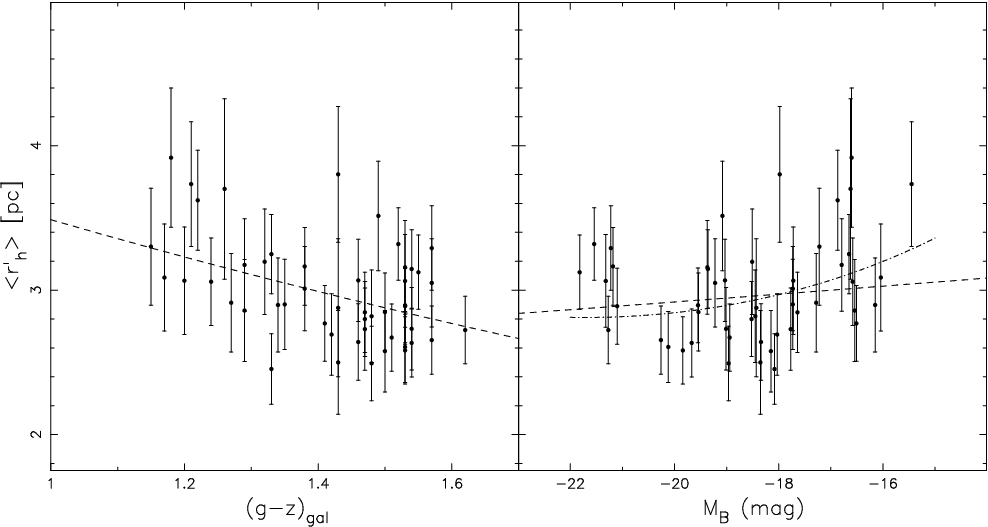}
\caption[]{({\it Left}) Mean GC half-light radius, $\langle r'_h \rangle$,
corrected to an underlying surface brightness of
$\mu_z = 21$ mag~arcsec$^{-2}$, plotted against
galaxy color, $(g-z)_{\rm gal}$. The dashed line is the 
least-squares line of best fit, showing a residual
dependence of $\langle r'_h \rangle$ on $(g-z)_{\rm gal}$.
({\it Right}) $\langle r'_h \rangle$ 
as a function
of galaxy absolute blue magnitude, $M_B$. 
The dashed line is the least-squares linear fit to the data. The
dot-dashed curve is the relation obtained
by folding the linear relation seen in the left panel
with a quadratic relation between $(g-z)_{\rm gal}$ 
and $M_B$ inferred from our sample. 
\label{fig:rhvsgzMB}
}
\end{figure}

Having examined the internal variation in $r_h$, we now turn our attention
to global dependencies. In this case, our sample is again restricted
in GC color, and the half-light radii have been corrected as given in
Equation~\ref{eq:rprime}: i.e., we are working with $r'_h$ values. Any
remaining dependency therefore must arise from global factors.
In Figure~\ref{fig:rhvsgzMB}, we show the
average half light radius $\langle r'_h \rangle$ plotted against
galaxy color $(g-z)_{\rm gal}$ and galaxy blue luminosity $M_B$
(left and right panels, respectively). 
In both cases, there may be  
residual trends, particularly among the bluer and fainter
galaxies, which seem to have somewhat larger $\langle r'_h \rangle$.
Linear fits of $\log \langle r'_h \rangle$ with galaxy color
and absolute magnitude give

\begin{equation}
\log \langle r'_h \rangle = (0.459 \pm  0.008) - (0.167 \pm 0.054)[(g-z)_{\rm gal}-1.5]
\end{equation}

\noindent and 

\begin{equation}
\log \langle r'_h \rangle = (0.465 \pm 0.010) + (0.004 \pm 0.005)(M_B+20).
\end{equation}

\begin{figure}
\plotone{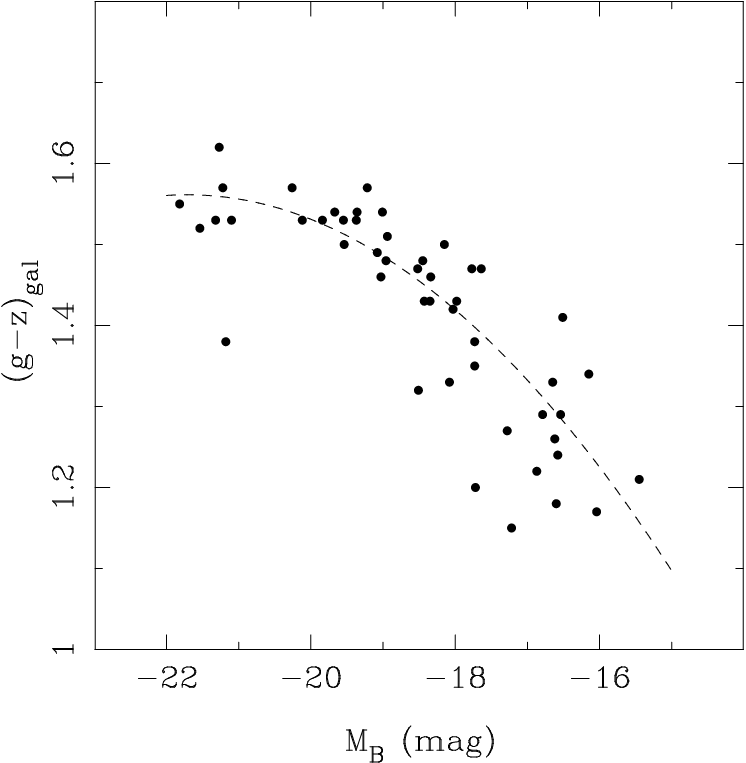}
\caption[]{Galaxy color, $(g-z)_{\rm gal}$, versus absolute blue
magnitude, $M_B$, for galaxies in our sample. The dashed line shows
the best-fit quadratic relation in the range $-22 \lae M_B \lae -15$.
\label{fig:gzvsMB}
}
\end{figure}

The coefficient of $M_B$ is not statistically significant, while the color
coefficient is significant at a $3\sigma$ level. Due to the well-known 
correlation between galaxy color and luminosity (Baum 1959; van den Bergh 1975),
the use of either $M_B$ or $(g-z)_{\rm gal}$ should be roughly equivalent 
as a tracer of the global variation in $r'_h$. For the ACSVCS sample, we can
determine the relationship between these parameters explicitly:
Figure~\ref{fig:gzvsMB} shows $(g-z)_{\rm gal}$ as a function of $M_B$,
along with the best-fit quadratic relation
 
\begin{equation}
(g-z)_{\rm gal} =  - 3.2 - 0.45 M_B - 0.010 M_B^2,
\end{equation}

\noindent which is valid in the range
$-22 \lae M_B \lae -15$. The need for 
a quadratic term is a reflection of the fact that the relation
flattens out at the high galaxy luminosities, with a plateau in
color after a certain  luminosity threshold is crossed
(cf. Tremonti et~al.\ 2004). This might provide a clue as to
why $\log \langle r'_h \rangle$ shows little or no dependence
on an $M_B$ 
term as compared with $(g-z)_{\rm gal}$. 
In any case, due to the better behavior and the added benefit
of being independent of distance, we will choose to model this
mild dependence using the galaxy color as opposed to 
its blue luminosity.

We now use the dependence on galaxy color to further correct the 
half-light radii:

\begin{equation}
r''_h \equiv r_h 10^{-0.016(\mu_z-21) + 0.17[(g-z)_{\rm gal} - 1.5]}.
\label{eq:rprime2}
\end{equation}

\subsection{Dependence on GC Color}

Using the half-light radii corrected according to Equation~\ref{eq:rprime2},
we now return to the question of the color dependence of half-light radii.
In doing so, we drop the restriction to blue GCs imposed in
\S~\ref{sec:dep:int}.

\begin{figure}
%\epsscale{0.85}
\plotone{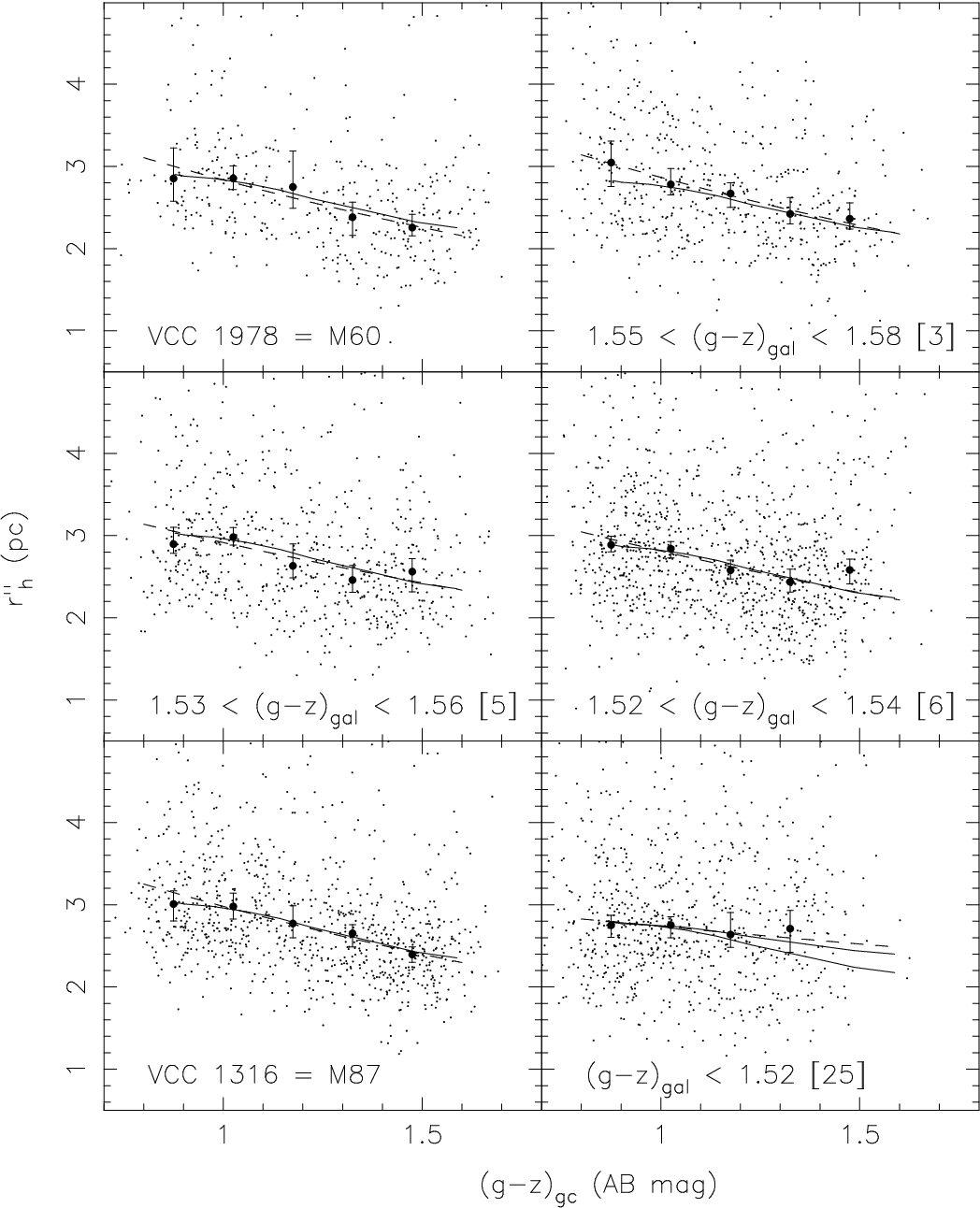}
\caption[]{
GC half-light radii, $r''_h$, corrected for the effects
of surface brightness and galaxy color, plotted against GC
color for GC subsamples formed by combining galaxies in 
color bins as described in the text. The
color bins, or the name of the individual galaxies
if only one present, are recorded in the different
panels. The number of galaxies
used to create the GC subsamples is indicated in brackets.
The large symbols indicate the median $r''_h$
in different GC color bins. The error bars indicate the 
99\% confidence intervals. The dashed line represents
the best linear fit of the form $\log r''_h \propto (g-z)_{\rm gc}$
to the data (transformed to $r''_h$). 
The solid line represents the prediction
of the behavior of $r''_h$ with $(g-z)_{\rm gc}$ from a model
from Jord\'an (2004) with a central potential $W_0 = 9$.
For the bin with $(g-z)_{\rm gal} < 1.52$ we have included
a second model with $W_0 = 7$ (the upper one).
\label{fig:gzrh}
}
\end{figure}

To quantify variations of $r''_h$ with color, we must have enough GCs
in any given color bin to get an accurate measure of the mean
behavior. This requirement imposes a limitation on the luminosity
of the galaxies to be studied, as less luminous galaxies
have GC systems which are almost exclusively metal-poor (i.e., 
many of the dwarfs have only a handful of GCs with
$(g-z)_{\rm gc} \gae 1.2$). Thus, we have decided to create composite
GC sets by combining galaxies in color bins
created by demanding at least 100 GCs with colors
in the range $1.3 \le (g-z)_{\rm gc} \le 1.5$. In addition, we
consider only those galaxies which have a number of GCs in the range
$1.3 < (g-z)_{\rm gc} < 1.6$ that is at least 10\% of the number found
in the range $0.8 < (g-z)_{\rm gc} < 1.1$. The motivation for this
condition is that adding galaxies which have purely metal-poor GC systems
adds no information to the characterization of the $r''_h$ color
dependence; indeed, because of uncertainties in the measured distances,
such galaxies would add scatter only in
the blue population while contributing few or no red GCs.
Figure~\ref{fig:gzrh} shows the resulting GC samples, which in some cases 
consist of a single galaxy. Also shown are the best-fit linear relations,
$\log r''_h = a(g-z)_{\rm gc} + b$ (transformed to $r''_h$), 
the median values of $r''_h$ in bins
of $(g-z)$ along with their 99\% confidence intervals, and 
a model from Jord\'an (2004) with central
potential\footnote{Jord\'an (2004) modelled the effects
of mass segregation using multimass isotropic
Michie-King models (Gunn \& Griffin 1979), which are characterized
by their central potential $W_0$, which is the gravitational potential
at the center of the cluster normalized by a characteristic velocity
variance. A value of $W_0=9$ is comparable to a value of $c\sim 1.6$
for the concentration of a single-mass isotropic King (1966) model, 
with the concentration increasing with $W_0$.} 
$W_0=9$, normalized such that the model at [Fe/H]$=-1.5$ coincides
with the median $r''_h$ at $(g-z)_{\rm gc} = 1.05$ mag.

It is apparent that the $r''_h$ show a significant correlation
with $(g-z)_{\rm gc}$ for all of the subsamples in Figure~\ref{fig:gzrh}.
Except for the group of bluer galaxies shown in the bottom left panel, 
the behavior is very consistent, with

\begin{equation}
\log r''_h \propto (-0.17{\pm}0.02) (g-z)_{\rm gc},
\end{equation}

\noindent valid
in the range $0.8 \lae (g-z)_{\rm gc} \lae 1.6$. The bluer
bin shows a somewhat shallower behavior, with a slope of $-0.09 \pm 0.04$, but
still within $\sim 2\sigma$ of the mean behavior of the redder
galaxies. Because the bluer bin required 25 galaxies for 
its construction, it is much more prone to 
uncertainties arising from errors in the measured
distances. Futhermore, these galaxies contain few red GCs, so their
usefulness in probing
the relation between $r''_h$ and $(g-z)_{\rm gc}$ is limited.
We will therefore assume that the behavior
exhibited by galaxies with $(g-z)_{\rm gal} \gae 1.52$ is valid
universally, and will use this relation to perform the final correction
of the GC half-light radii (see below).

It is remarkable how well a single model from Jord\'an (2004),
with $W_0=9$, does in matching the data in Figure~\ref{fig:gzrh}.
With the possible exception of the bin containing the bluest
galaxies, it reproduces the observed trends very well.
As mentioned in \S\ref{sec:intro}, only two mechanisms have been
proposed to explain the observed tendency for red GCs to be smaller
than their blue counterparts. Larsen \& Brodie (2003) proposed
that this effect could be the consequence of projection effects, combined
with the differing spatial distributions of the red and blue GCs. The
red GCs are observed to be more centrally concentrated so that, if
the GCs follow a relationship between
galactocentric distance and $\langle r_h \rangle$ which is similar
to that of GCs in the Milky Way, projection effects can lead to the
red GCs appearing systematically smaller (due to the fact that they
are, on average, physically closer to the galaxy center). As noted
by Larsen \& Brodie (2003), however, this mechanism requires some
fine-tuning to work. Jord\'an (2004), on the other hand, has proposed 
that the observed size difference can be explained as the consequence
of mass segregation within individual GCs, combined with the dependence
of stellar lifetimes on metallicity. Simply stated, if the half-{\it mass}
radii are assumed to be independent of metallicity, the half-{\it light}
radii will, at fixed age, differ because the brightest stars, which
dominate the overall light profile, will be more massive in red 
(metal-rich) GCs and thus more centrally concentrated.

We can use our new measurements to discriminate between these alternatives. 
First,
we note that our observations point to a significantly shallower
relation between $r_h$ and galactocentric radius in early-type galaxies
than the one observed in our Galaxy. This latter relation was assumed by 
Larsen \& Brodie (2003) and is necessary for the projection mechanism to 
explain the observed trend. Using 
shallower relations they found that projection effects
are unable to reproduce the observed size difference.
Thus, our measurements seem to rule out projection effects as the primary
mechanism responsible for the observed difference, although it is
certainly possible that they play some part in its origin.

In addition, the importance of projection effects is expected to depend on
the fraction of the galaxy surveyed, as they are expected to decrease 
strongly beyond one effective radius of the GC system, and to 
disappear entirely as $r_p \rightarrow \infty$ (Larsen \& Brodie 2003).
The galaxies with $(g-z)_{\rm gal} \gae 1.52$ span an enormous
range in $r_{\rm lim}/r_e$, where $r_{\rm lim}$ is the upper
radii for inclusion of GCs in each galaxy and $r_e$ is the 
effective radius of the galaxy light. 
These galaxies have $1 \lae r_{\rm lim}/r_e \lae 20$, with 
a median of $r_{\rm lim}/r_e = 3.3$ and an interquartile range
of $\sim 8$. That the observed difference in $r_h$
is observed to be nearly identical among these galaxies argues
strongly for a mechanism that is independent of projection effects.
It might be argued that the fact that the decrease in size seems to be milder 
for the bluest galaxies is consistent 
with at least some variation due to projection
effects. But the distribution of $r_{\rm lim}/r_e$ for the galaxies
in the bluest bin is not extremely different from those in the 
redder ones, having $2 \lae r_{\rm lim}/r_e \lae 25$, a median
of $r_{\rm lim}/r_e = 4.8$ and an interquartile range of $5.5$.
Thus, it would be fortuitous that the change of regime
occurs between samples with median $r_{\rm lim}/r_e$ of
$\sim 3.3$ and $\sim 4.8$. 

The mechanism of Jord\'an (2004) seems to fare better,
despite the fact that it is a clear oversimplification to compare the data
with a single model with $W_0=9$. In reality, GCs will  span a
range in $W_0$, and there could well exist a systematic difference
in the average central potential of GCs belonging to the blue 
(low-luminosity) galaxies. As we show in Figure~\ref{fig:gzrh}, a model
with $W_0=7$ for the bluer bin is able to reproduce the
observations quite well. In any case, the overall conclusion to
be drawn from the data in Figure~\ref{fig:gzrh} is that,
whatever is responsible for the size difference between
red and blue GCs does, this difference does not seem to 
arise mainly from projection
effects but is instead the result of some property intrinsic to
the GCs.

\subsection{Combined results}

Having systematically explored the variations in half-light radii
with a number of internal, local, and global factors, we now use the
observed dependencies to correct the individual $r_h$ measurements to
the expected value for a GC with color $(g-z)_{\rm gc} = 1.2$~mag, 
at an underlying surface brightness of $\mu_z=21$ mag~arecsec$^{-2}$, 
in a galaxy with $(g-z)_{\rm gal}=1.5$ mag. 
In other words, we assume that only the mean of the 
$r_h$ distribution is changing to the above scaling relations,
so that the corrected radii, $\hat{r}_h$, are given by

\begin{equation}
\hat{r}_h \equiv r_h 10^{-0.016(\mu_z-21) + 0.17[(g-z)_{\rm gal} - 1.5]
+0.17[(g-z)-1.2]}.
\label{eq1}
\end{equation}

\begin{figure}
\plotone{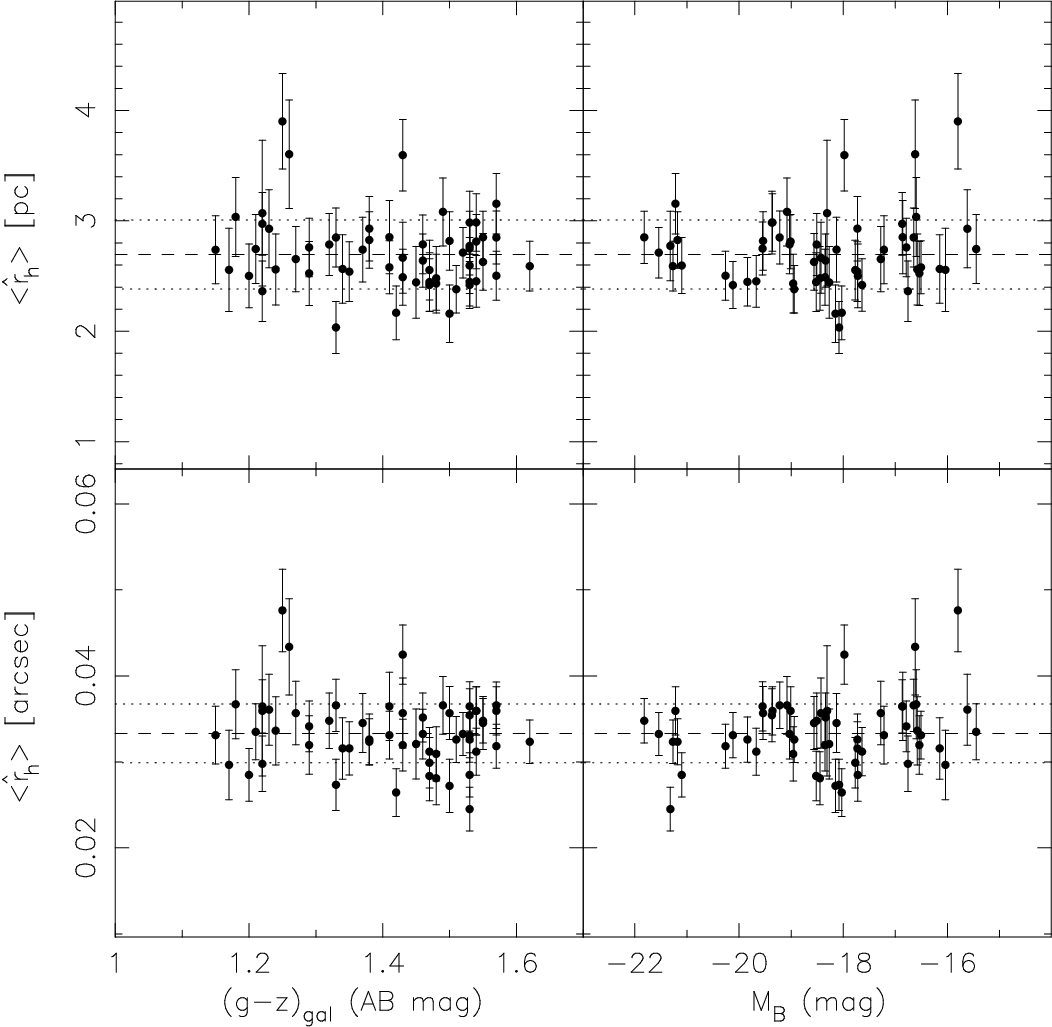}
\caption[]{ ({\it Upper Panels}) Fully corrected mean GC half-light
radius, $\langle\hat{r}_h\rangle$, in parsecs, plotted as a function
of galaxy color, $(g-z)_{\rm gal}$, in the left panel, and against
and absolute blue magnitude, $M_B$, in the right panel. The dashed
line indicates the biweight estimate of the mean,
$\langle\hat{r}_h\rangle = 2.72$ pc. The dotted lines
indicate the dispersion of 0.35 pc.
({\it Lower Panels})
Fully corrected mean GC half-light
radius, $\langle\hat{r}_h\rangle$, in arcseconds, plotted as a function
of galaxy color, $(g-z)_{\rm gal}$, in the left panel, and against
and absolute blue magnitude, $M_B$, in the right panel. The dashed
line indicates the biweight estimate of the mean,
$\langle\hat{r}_h\rangle = 0\farcs033$. The dotted lines
indicate the dispersion of $0\farcs0035$. VCC731, a member of the
W Cloud, has been excluded from the computation of these quantities.
\label{fig:rhcor}
}
\end{figure}

We plot $\langle\hat{r}_h\rangle$ for our sample galaxies, in both 
arcseconds and parsecs, against $(g-z)_{\rm gc}$ and 
$M_B$ in Figure~\ref{fig:rhcor}. It is clear that
$\langle\hat{r}_h\rangle$ is remarkably constant across our
sample galaxies. As such, this quantity appears to be a ``standard
ruler", suitable for use as an extragalactic distance indicator. We shall
explore this issue in more detail in \S\ref{sec:distance}.

We emphasize that none of the dependencies discussed above are likely
to be due to selection effects. Compact clusters at large galactocentric 
radii (or, equivalently, at low surface brightness) would be readily
detected and measured in our survey. The inner radial cut imposed
on each galaxy ensures that we are highly complete for extended
GCs. The decrease in
$\langle r'_h \rangle$ with galaxy color could be 
suspected to be due to contamination
preferentially affecting the small GC systems of faint, blue galaxies. However,
we have accurate estimates of the expected contamination measured
directly from our control fields (Peng et~al. 2005, in preparation)
so this is unlikely to be the case.

While from an observational point of view, the local and global factors can be
separated naturally, they could in principle arise from a single relation
which determines $r_h$. As a simple illustration, let us assume that
$r_h$ is related to the surface mass density, $\Sigma_{\rm gal}$, through the
relation
$\langle r_h \rangle \propto \Sigma_{\rm gal}^{-x} \equiv [(M_{\rm gal}/L_{\rm gal}) I_{\rm gal}]^{-x}$,
where $I_{\rm gal}$ denotes the galaxy surface brightness and
$(M_{\rm gal}/L_{\rm gal})$ its mass-to-light ratio. Clearly, in this case

\begin{equation}
\log \langle r_h \rangle \propto -x\log (M_{\rm gal}/L_{\rm gal}) + 0.4x\mu_z.
\end{equation}

\noindent The mass-to-light ratio will vary systematically with both 
the galaxy luminosity and color. Taking $(M_{\rm gal}/L_{\rm gal}) \propto L_{\rm gal}^{0.3}$ 
from Cappellari et~al. (2005), we get

\begin{equation}
\log \langle r_h \rangle \propto 0.12x M_B + 0.4x\mu_z.
\end{equation}

\noindent where we have used the fact that 
$L_{\rm gal} \propto 10^{-0.4M_B}$. Under 
this simple hypothesis, a value of $x\sim 0.04$ would be a 
reasonable description of the observations (cf. eqs 4 and 8). 
This simple example serves to show that empirical trends 
can guide our theoretical understanding of both GC formation
and galactic structure.

We will now explore the implications of the observed global behavior
for models which make quantitative predictions regarding the half-light
radii of GCs and their dependence on environment.

\section{Theoretical Considerations}
\label{sec:th}

Our finding that the half-light radii of GCs show clear correlations
with some properties of their host galaxies 
should not obscure the remarkable fact that 
over a factor of $\sim 350$ in 
galaxy luminosity, $\langle r_h \rangle$ varies by only $\sim 10 \%$ 
(after accounting for internal and local factors). Clearly, one 
could assume that, to first
order, $\langle r_h \rangle$ is constant: 
$\langle r_h \rangle \sim L_{\rm gal}^0$.
Thus, our results demonstrate that any paradigm for the formation of
GCs should predict an exponent $\eta$ in the relation
$\langle r_h \rangle \sim L_{\rm gal}^\eta$ which is very close to zero.
In this section, we show that two possible hypothesis for the 
origin of the observed $r_h$ of GCs do lead to
roughly this result. It should be stressed at the onset, however,
that the level of theoretical understanding we have
on process of GC formation is sketchy at best. Indeed, the observations
are far ahead, in the sense that they provide an
increasingly clear, albeit sometimes puzzling, set of 
constraints to which any theory of GC formation must conform.

\subsection{Half-Light Radii as the Result of Pressure-Confined Proto-GC Clouds}

It has been suggested that the half-light radii of GCs reflect the
structure of self gravitating proto-GC clouds which, at the time
of GC formation, were
in pressure equilibrium with the surrounding gas. The
structure of such clouds has been calculated by McLaughlin \& Pudritz 
(1996), who determine that their limiting radius
satisfies the relation $R_{\rm lim} \propto M^{0.5} P_{gas}^{-1/4}$ 
(cf. equation 2.11 in McLaughlin 
\& Pudritz 1996), where $P_{gas}$ is the pressure of the ambient gas. 
If we posit that the average $r_h$ observed today is linearly related
to $R_{\rm lim}$, then it follows that 
$\langle r_h \rangle \propto M^{0.5} P_{gas}^{-1/4}$, where we assume
that GCs of differing masses that form at approximately
the same radius in a galaxy come from clouds with differing velocity dispersions.
It is immediately apparent that this hypothesis is in strong
conflict with observations due to the factor $M^{0.5}$. The observational
evidence presented in \S3 argues
for $\langle r_h \rangle \propto M^0$. To resolve this discrepancy
one might propose that the star formation efficiency
scales as some power of the proto-GC cloud binding energy 
(e.g., Ashman \& Zepf 2001). While such a
scaling relation could, in principle, erase the mass dependence, there is no 
physical basis to justify such a scaling.
In any case, we will carry on by supposing that a mechanism 
exists to render this pressure support picture viable, and examine
what this scenario predicts for the global behavior of 
$\langle r_h \rangle$ with host galaxy luminosity. 

We begin by supposing
that $\langle r_h \rangle \propto P_{gas}^{-1/4}$. If
the ambient pressure is distributed like the dark matter,
then averaging over an entire galaxy gives

\begin{equation}
\langle r_h \rangle \propto P_{\rm ave}^{-1/4} \propto 
\left(\frac{M^2_{\rm vir}}{r^4_{\rm vir}}\right)^{-1/4}
\end{equation}

\noindent where $M_{\rm vir}$ and $r_{\rm vir}$ are the virial
mass and virial radius of the dark halo.  We assume now that
early-type galaxies all formed at similar large redshifts.
The halos should then have the same average density,
or $M_{\rm vir} \propto r_{\rm vir}^3$.
Since the circular velocity at the virial radius is
given by $v_{\rm vir}^2 \equiv G M_{\rm vir}/ r_{\rm vir}$, we have
then that $r_{\rm vir} \propto v_{\rm vir}$ and therefore we expect 
$\langle r_h \rangle \propto v_{\rm vir}^{-1/2}$.

In order to progress further, we assume now that the
dark matter halos are described by NFW profiles 
(Navarro, Frenk \& White 1995). Then
the circular velocity $V_c$ profile is 

\begin{equation}
\left[\frac{V_c(r)}{v_{\rm vir}}\right]^2 = 
\frac{1}{x}\frac{\ln (1+cx)-cx/(1+cx)}{\ln (1+c) - c/(1+c)},
\end{equation}

\noindent where $x \equiv r/r_{\rm vir}$ and $c$
is the concentration parameter. This velocity
profile implies a peak circular velocity at radius
$cx \simeq 2.163$, at which point the value is

\begin{equation}
\left(\frac{V_p}{v_{\rm vir}}\right)^2 \simeq 
0.2162\frac{c}{\ln (1+c)-c/(1+c)} \approx 0.2162[2.742c^{0.4}]
\end{equation}

\noindent where the second approximation is accurate to 
within $\sim 5\%$ in the range $4\leq c \leq 20$.
Furthermore, simulations show (Bullock et~al.\ 2001) that
$c \sim M_{\rm vir}^{-0.13}$ and thus 
$V_p^2 \sim c^{0.4}v_{\rm vir}^2 \sim v_{\rm vir}^{1.84}$.
Therefore, $\langle r_h \rangle$ is related to the peak circular
velocity of the halo through

\begin{equation}
\langle r_h \rangle \propto v_{\rm vir}^{-1/2} \sim V_p^{-0.54}
\end{equation}

Finally, Ferrarese (2002) connects $V_p$ to the {\it stellar} 
velocity dispersion of the central bulge by 

\begin{equation}
V_p \sim \sigma_0^x, \,\,\,\, x=0.89\pm0.04
\end{equation}

\noindent and the Faber-Jackson relation has 
$\sigma_0 \propto L_{\rm gal}^{1/4}$
so ultimately

\begin{equation}
\langle r_h \rangle \propto L_{\rm gal}^{-0.12}.
\label{eq:press}
\end{equation}

Note that a scaling relation of the form of Equation~\ref{eq:press}
can also be obtained much more simply
by assuming that $\sigma_0 \propto v_{\rm vir}$, and as
consequence $v_{\rm vir} \propto L_{\rm gal}^{1/4}$ and
$\langle r_h \rangle \propto v_{\rm vir}^{-1/2} \propto L_{\rm gal}^{-1/8}$.
In both cases, we get a very mild dependence of $\langle r_h \rangle$
with $L_{\rm gal}$, albeit stronger than that which is observed.
We note that if we had assumed that the average ambient 
pressure of the proto-GC clouds comes from gas distributed as the
present-day starlight and used fundamental plane scaling relations,
we would get a similar dependence but of opposite sign. 
That is to say, in this case we would posit 

\begin{equation}
\langle r_h \rangle \propto P_{\rm ave}^{-1/4} \propto 
\left(\frac{M_{\rm gal}^2}{r_e^4}\right)^{-1/4}
\end{equation}

\noindent where $r_e$ is the effective radius of the 
galaxy light. The galactic fundamental plane scalings 
of $r_e \propto L_{\rm gal}^{0.8}$ and $M/L \propto L_{\rm gal}^{0.3}$ 
(Ha\c{s}egan et~al. 2005 and references therein) then give
$\langle r_h \rangle \propto L_{\rm gal}^{0.15}$. Overall, it 
seems fair to conclude that the pressure hypothesis
predicts roughly $\langle r_h \rangle \propto L_{\rm gal}^0$,
in accord with the observations from \S3.

\subsection{Half-Light Radii as a Consequence of Tidal Limitation}

An alternative hypothesis is that the half-light radii of GCs
are proportional to their tidal radii, $r_t$. 
This runs immediately into the problem of the non-homologous
nature of GCs. In fact, the
King concentration parameter $c$ is correlated with the 
GC mass through the rough relation $10^c\sim M^{0.4}$, 
which could be the reflection of an original ``fundamental
line'' (McLaughlin 2000). The $r_h$ in a King model
is a function of $c$ times $r_t$ so that, via the
dependence of $c$ on $M$, it follows that a simple 
proportionality between $r_h$ and $r_t$ for all GCs 
cannot be appropriate.
Also, there is no clear mechanism to physically transmit 
the information from the tidal regions to determine the
half-light radius.
Finally, if the $r_t$ are determined by the tidal field of the host galaxy,
then we would have 
$\langle r_h \rangle \propto \langle r_t \rangle \propto M^{1/3}$ 
(Binney \& Tremaine 1987). 
Thus, we would need to identify a mechanism that erases the mass 
dependence as in the pressure hypothesis above.
 
Assuming that such a mechanism exists,
what would this hypothesis predict for the global behavior? In this case 
$\langle r_h \rangle$ should be related to the mean density of the galaxy's
dark matter halo, or 
\begin{equation}
\langle r_h \rangle \propto (M_{\rm vir}/r_{\rm vir}^3)^{-1/3}
= \mbox{constant} \sim L_{\rm gal}^0
\end{equation}
Therefore, this picture would also
predict a roughly constant $\langle r_h \rangle$ across galaxies.
It is interesting
that both hypotheses predict roughly the observed global behavior.
We conclude that the current observations cannot
discriminate between these two scenarios.
Indeed, the observational constraints seem to be ahead of what our
limited knowledge of GC formation can predict, although it is
clear that fundamental aspects of GC formation are encoded in them.

\section{Half-Light Radii as a Distance Indicator}
\label{sec:distance}

\begin{figure}
%\epsscale{0.5}
\plotone{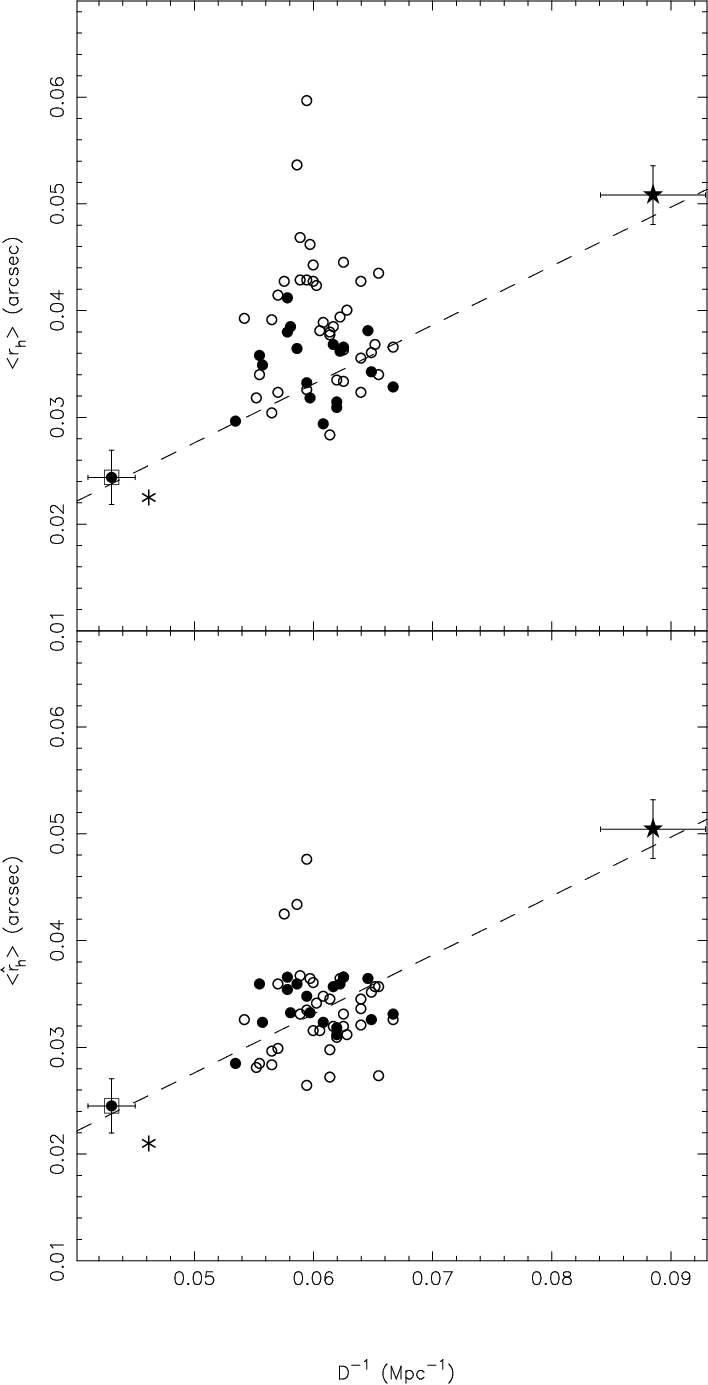}
\caption[]{({\it Upper panel}) $\langle r_h \rangle$ (no corrections)
versus inverse distance $D^{-1}$. Filled symbols are galaxies 
with $M_B < - 19$ mag, while open symbols are those with $M_B \ge -19$ mag.
The star is NGC~4697 and the filled circle surrounded by a square is VCC~731.
The dashed line represents the relation $\langle r_h \rangle \propto D^{-1}$
normalized to be $0\farcs033$ at $D=16.5$ Mpc, and it is {\it not} a fit.
Error bars are included only for NGC~4697 and VCC~731 for clarity. The asterisk
is VCC~575.
({\it Bottom panel})$\langle \hat{r}_h \rangle$ versus 
inverse distance $D^{-1}$. Filled symbols are galaxies 
with $M_B < - 19$ mag, while open symbols are those with $M_B \ge -19$ mag.
The star is NGC~4697 and the filled circle surrounded by a square is VCC~731.
The dashed line represents the relation $\langle r_h \rangle \propto D^{-1}$
normalized to be $0\farcs033$ at $D=16.5$ Mpc, and it is {\it not} a fit.
Error bars are included only for NGC~4697 and VCC~731 for clarity. The asterisk
is VCC~575. 
}
\label{fig:distance}
\end{figure}

In the upper panel of 
Figure~\ref{fig:distance}, we plot $\langle r_h \rangle$ in arcseconds
versus inverse distance, $D$, for those galaxies in our sample with
SBF distance measurements (Mei et~al. 2005c). 
If $\langle r_h \rangle$ is constant in physical units, then we expect
$(\langle r_h \rangle/1\arcsec) \propto D^{-1}$ at low redshift 
(the dashed line).
Although the vast majority of our sample galaxies 
are located at very nearly the same
distance (i.e., the mean Virgo Cluster distance of $D = 16.5\pm 0.25$ Mpc;
Mei et~al. 2005c), two galaxies in our sample give us some additional
leverage to test this idea: (1) VCC~731 (NGC~4365) is a member of the 
Virgo W Cloud  (de Vaucouleurs 1961) and has an 
SBF distance of $D=23.3\pm$0.6 Mpc 
(Mei et~al. 2005c); and (2) NGC~4697, a non-member of the Virgo Cluster
which lies in the foreground, with an SBF distance
of 11.3$\pm$0.5 Mpc (Tonry et~al. 2001; Mei et~al. 2005c).\footnote{Tonry
et~al. (2001) list the distance as 11.7 Mpc. To compare this value to our
new SBF distance, the zeropoint must be adjusted to match that
used in Mei et~al. (2005b). As a consequence, the distance modulus
in Tonry et~al. (2001) should be shifted by --0.06 mag, so that
$D = 11.4\pm1.6$ Mpc. This is in excellent agreement with our new
measurement of 11.3 Mpc.}

The upper panel of Figure~\ref{fig:distance} reveals a scaling of
$\langle r_h \rangle$ with $D^{-1}$ which has roughly the expected form,
albeit with appreciable scatter for some of the fainter Virgo members. 
Note, however, that
the half-light radii plotted in this panel have not been corrected for any
dependence on GC or host galaxy properties as discussed in \S3. In the bottom
panel of this figure, we plot the average corrected half-light radii,
$\langle \hat{r}_h \rangle$, against inverse distance. 
The dashed line, $\langle \hat{r}_h \rangle \propto D^{-1}$, is drawn so
that $\langle \hat{r}_h \rangle$ has a value of $0\farcs033$ at our adopted mean
distance of Virgo (Mei et~al. 2005b).  Note that
line is {\it not} a fit, although it accurately predicts the measured 
$\langle \hat{r}_h \rangle$ for both VCC~731 and NGC~4697. 
It is clear that $\langle \hat{r}_h \rangle$ has the 
potential to be a powerful standard ruler 
for distance estimation. But as the upper panel of 
Figure~\ref{fig:distance} shows, even without any correction
the directly measured $\langle r_h \rangle$ can be a useful distance
indicator when restricted to the luminous galaxies 
(a class to which both
VCC~731 and NGC~4697 belong). And trading some accuracy for simplicity, 
the mild nature of the required corrections means that a directly 
measured $\langle r_h \rangle$ can give a rough distance measurement
for any early-type galaxy.

It should be noted that at least two other galaxies in the 
ACSVCS have celestial positions close to the W Cloud: VCC~571 and 
VCC~575. While they do not have enough GCs remaining after the selection 
process described in \S~\ref{sec:obs} to be included in our sample 
(they are left with just 3 GCs each), 
their measured $\langle \hat{r}_h \rangle$
are nevertheless consistent with membership in the W Cloud. More
specifically, VCC~571 has $\langle \hat{r}_h \rangle=0\farcs023$, while
VCC~575 has $\langle \hat{r}_h \rangle=0\farcs021$. An SBF distance
is available only for VCC~575, which confirms that it is at a
similar distance as VCC~731. While the number of GCs used to 
measure $\langle \hat{r}_h \rangle$ for VCC~575 is indeed low, 
it is reassuring that its measured value is consistent with it being
at the distance of the W Cloud ($\approx 23$ Mpc), 
lending additional support to the validity of
using $\langle \hat{r}_h \rangle$ as a distance indicator.
The position in the sky and the measured $\langle \hat{r}_h \rangle$
for VCC~571 suggest that it is indeed a member of the W Cloud.

We now put forward the following method and calibration for the use of
GC half-light radii in distance estimation. 
If a set
of $\{r_{h,i}\}$ are measured on a given galaxy they should first
be individually 
corrected for their color, underlying surface brightness and the host
galaxy color by computing
$\hat{r}_{h,i}$ according to equation~\ref{eq1}.  The surface brightness
can be measured from the same frame where the GCs
are detected and analyzed, and all of the quantities
needed to perform the correction are distance independent.
Then, if the median of the $\hat{r}_{h,i}$ is 
$\langle \hat{r}_h \rangle$ in arcseconds, the distance in Mpc 
follows from  

\begin{equation}
D = (0.552 \pm 0.058)\left( \frac{ \langle {\hat r}_h \rangle }{ 1\arcsec }
		\right)^{-1} \, {\rm Mpc} \, 
\label{eq2}
\end{equation}

\noindent where the uncertainty reflects the uncertainty in the 
assumed mean Virgo distance and the PSF systematics. 
While the expression above is in principle applicable to any 
early-type galaxy within the color/luminosity range spanned by our sample 
galaxies [i.e., $1.1 \lae (g-z)_{\rm gal} \lae 1.6$ and
$-22 \lae M_B \lae -15.5$], in practice this technique
would be applied almost exclusively to luminous giants which
have populous GC systems, and where large samples of GCs 
can be easily accumulated and contamination
is less of an issue. 
Additionally, in equation~\ref{eq1} measurements of the $(g-z)$ 
colors for the GC and galaxy are required, which would entail
observing in these same bandpasses or transforming the color in 
hand to $(g-z)$, which would add some uncertainty. Needless to
say, this also means that observations in at least two 
bands are needed to perform the
correction. Thus, we note that for luminous galaxies (i.e., those
with $M_B \sim -21$ mag)
a distance may be obtained without the benefit of color
information by correcting the set of $r_{h,i}$ with the following
approximation to equation~\ref{eq1}:

\begin{equation}
\hat{r}_h \simeq r_h 10^{-0.016(\mu_z-21) + 0.01}.
\end{equation}

\noindent The distance then follows by applying equation~\ref{eq2}. The
additional uncertainty incurred by this approximation if applied
to a giant elliptical can be estimated to be of order a few percent.
Of course $M_B$ won't be known beforehand and it would not be possible to know
if the approximation is valid. In practice, this complication can be circumvented
by applying the method only to the $\sim 10$ brightest early-type members of
a given galaxy cluster.

What accuracy can be expected from this method?  The rms deviation from a constant
value $\langle r_h \rangle = 0\farcs0333$ among our Virgo galaxies is
$0\farcs0034$ (where VCC~731 has been excluded). Thus, a single measurement
can be expected to give a distance with external accuracy of $\approx 11\%$. 
It is interesting to see that using our sample 
of $\langle \hat{r}_h \rangle$ we can
already set an {\it upper} limit to the depth of Virgo 
(once again, excluding the background W Cloud). Assuming that
$\langle \hat{r}_h \rangle$ is strictly constant and neglecting
observational uncertainties, the observed dispersion translates into 
a distance dispersion of $\sigma_D \lae 1.7$ Mpc, or
a 1-$\sigma$ line-of-sight {\it depth} of 3.4~Mpc. While this finding
is in good agreement with several previous measurements of the depth of Virgo 
around M87 (e.g., West \& Blakeslee 2000,
Neilsen \& Tsvetanov 2000, Jerjen et~al.\ 2004), it
effectively rules out the claim of Young \& Currie (1995)
that the Virgo Cluster has a very elongated distribution 
(extending from $\sim$ 8-20 Mpc) along the line of sight.
Thus, this approach complements previous evidence against this claim from
SBF measurements (Jerjen et~al.\ 2004; Mei et~al. 2005c).

\subsection{Geometrical Calibration and an Independent Distance to Virgo}

Up to now we have relied on our SBF distance measurements to establish the physical
size of GCs in Virgo and to provide a calibration for the use of half-light radius
as a distance indicator.
Being reliant on SBF means that our distance measurement method
is tied to the primary calibrators of the SBF method itself, and
in particular to the calibration of the Cepheids.
In this section, we calibrate the method using the Milky Way GC system only,
thereby providing a simple geometric calibration of the method which is largely
independent of any other extragalactic distance measurement method.

First, we note that all of the dependencies used above
to compute corrected radii, $\hat{r}_{h,i}$, are independent
of our SBF distances.
Thus, we keep the form of these dependencies
intact and attempt derive the zeropoint for the physical size using
the Milky Way GC system. Of course, this approach faces has the difficulty
of estimating the parameters needed for the corrections for the Milky Way, 
and the complication that our Galaxy is a disk system as opposed to the
early-type galaxies targeted in the ACSVCS. These caveats notwithstanding, 
we now estimate a rough zeropoint to provide a direct calibration.

The quantities needed to determine $\langle \hat{r}_h \rangle$ for Galactic
GCs are the $(g-z)$ colors of the GCs and the Galaxy itself, and the surface
brightness profile in $z$. Instead of working with the latter, we choose
to use $\log (r_{\rm p}/r_e)$, where for the effective radius we adopt
the value given by de Vaucouleurs \& Pence (1978), $r_e \approx 5$ kpc. We 
then use the Harris (1996) catalog of Galactic GCs to obtain the $r_h$, 
[Fe/H] and galactocentric distances $R_{\rm gc}$.
In order to estimate their $(g-z)$ colors,
 which are not directly available, we transform from [Fe/H] to $(g-z)$ as in 
Jord\'an et~al. (2004b) (which assume an age of 13 Gyr). 
Finally, we take the integrated $(g-z)_{\rm gal}$ color
of our Galaxy to be $(g-z)_{\rm Sbc} = 1.17$, the value appropriate for
a type Sbc galaxy according to Fukugita et~al.\ (1995).
With these quantities in hand, we obtain a set of corrected half-light $\{\hat{r}_{h,i}\}$
radii using the prescriptions discussed above, namely 

\begin{equation}
\hat{r}_{h,i} = r_{h,i}10^{-0.07\log (r_{\rm p}/5) 
+ 0.17[(g-z) - 1.2) + 0.17(g-z)_{\rm Sbc} - 1.5]},
\end{equation}

\noindent for each of 100 simulated Galactic datasets with
randomly projected galactocentric distance 
and with GCs selected according to our selection procedures 
(cf. \S~\ref{subs:local}). 

From these simulations, we find $\langle \hat{r}_h \rangle \approx 2.54\pm0.1$~pc
for the Galactic GC system. Note that the quoted uncertainty
does not include systematics arising from our assumed
properties of the Galaxy. These can be roughly estimated to be 
at least $15\%$ by varying the assumed $r_e$ by 20\% and 
changing the assumed color to that of a galaxy of type Sab and Scd. 
Using the fact that  the observed $\langle \hat{r}_h \rangle$ 
for our sample is 
$0\farcs033$, we find that the mean distance of Virgo using
this direct calibration is 16$\pm$2.3 Mpc. 
While this calibration should be viewed with some caution
(being based on a late-type galaxy for which we do not have a direct
measurement of its global characteristics), it is comforting to see
that the distance of Virgo is in good agreement
with other methods. Moreover, because this estimate circumvents
several of the problematic issues involved in other methods used
to establish Virgo's distance, such as the metallicity dependence of
the Period-Luminosity relation for Cepheid variables and the distance modulus to
the Large Magellanic Cloud, it provides a largely independent
confirmation of the Cepheid/SBF distances.

\section{The Form of the $r_h$ Distribution}
\label{sec:rhdist}

So far we have examined only the dependencies and properties of a location
parameter of the distribution of GC half-light radii in early-type galaxies. 
In fact, in applying the various corrections to the set of $r_{h,i}$,
we have tacitly assumed that
only $\langle r_h \rangle$ was changing and that the {\it shape} of
the $r_h$ distribution is invariant. We now show that this assumption
is well justified and present an analytic
form which accurately describes the observed $r_h$ distributions. 
Throughout this section, we will work with the fully corrected 
half-light radii, $\hat{r}_h$ and we additionally include galaxies
that have less than 6 GCs after the cuts described in \S~\ref{sec:obs}.

\begin{figure}
\plotone{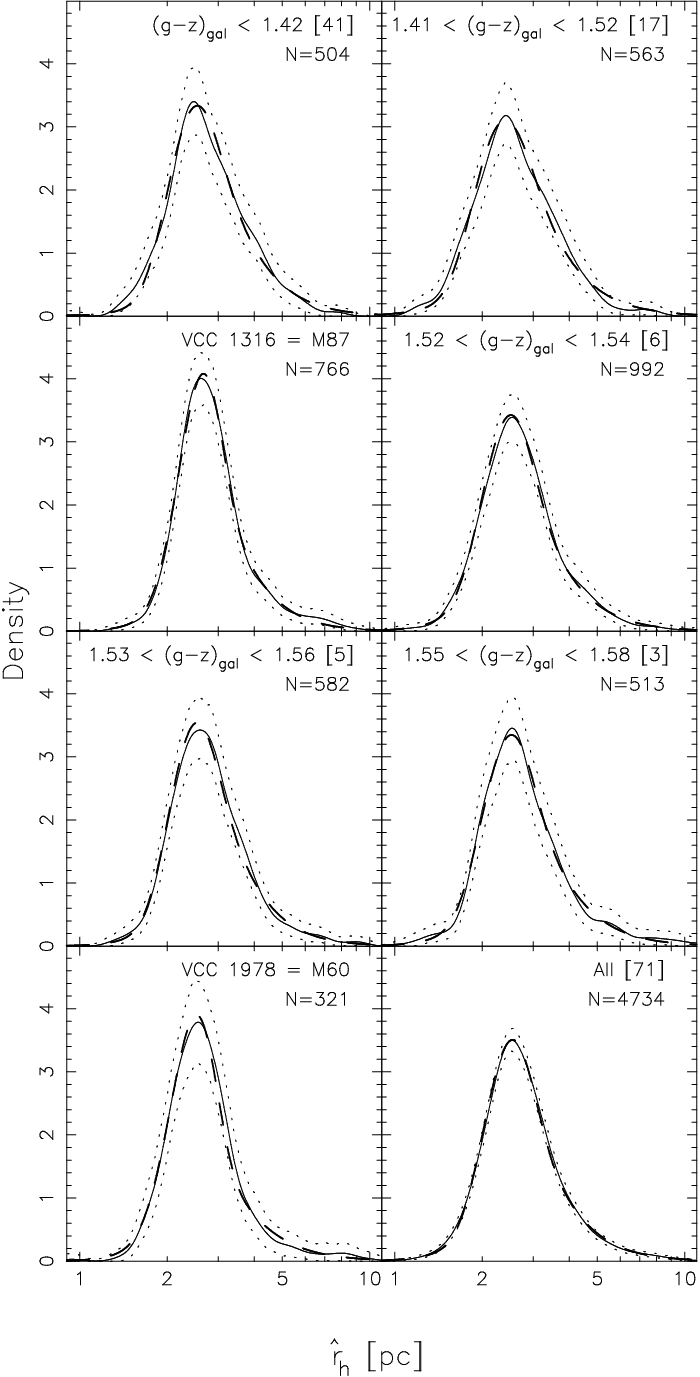}
\caption[]{ Distribution of $\log \hat{r}_h$ for GC
samples built by combining galaxies in 
galaxy color bins as described in the text.
The color bins (or the name of the individual galaxies
if only one is present) are indicated in the different
panels. The number of galaxies
used in creating the combined GC sample is indicated in brackets
and the number of GCs included is indicated by the value of $N$.
The bottom right panel is built by combining all galaxies in the sample.
The solid curves show Gaussian kernel density estimates
with a bandwidth $h=0.035$; the dotted curves show their 99\%
confidence bands. The thick dashed line represents the best-fit
function ${\cal J}$ (Equation~\ref{eq:J}) convolved with a
Gaussian with $\sigma=h$.
}
\label{fig:rhdist}
\end{figure}

To study the form of the distribution of half-light radii
across the color and luminosity range of our galaxies, 
we chose to create bins in galaxy color. To minimize statistical
noise, we require that each subsample contains at least 500 GCs (the reddest
galaxy, VCC~1978, is assigned  a bin of its own although it does not satisfy 
this condition).
The resulting $\hat{r_h}$ distributions are shown in Figure~\ref{fig:rhdist},
along with the distribution of the combined sample.
Each panel shows a Gaussian kernel density estimate of the observed
$\log \hat{r}_h$ distribution, with a bandwidth $h=0.035$ (solid curves).
The corresponding 99\% confidence bands are indicated by the dotted curves.
It is apparent
that the distribution of $\log r_h$ is very similar across the galaxy color
range of our sample. The primary features are a prominent peak
at $\hat{r_h} \sim 2.5$ pc and an extended, large-$\hat{r}_h$ tail 
which is much heavier than if the distribution was normal.
The distribution is noticeably skewed, with an excess on the right side
of the peak.

We emphasize that the distributions shown in Figure~\ref{fig:rhdist} are rather
uncertain for $\hat{r}_h \lae 1$~pc, as the
measurements in that regime begin to suffer from biases to slightly larger $r_h$,
due to the fact that the GCs are close to being unresolved.
This difficulty notwithstanding, the decline toward small $r_h$
is certainly real. Unfortunately, the
data do not allow us to characterize the regime $\hat{r}_h \lae 1$~pc accurately,
so the following discussion should be taken to apply 
to the range $1 \lae \hat{r}_h \lae 10$ pc only. Note that the upper radius is 
imposed by our sample selection; in practice, this limit is set by
the difficulty of distinguishing extended GCs from
background galaxies. Thus, highly extended GCs, which certainly exist
in our Galaxy, are not considered here.
In any case, there appears to be no sudden upturn in the distribution
for $\hat{r}_h \lae 1$~pc or $\hat{r}_h \gae 10$~pc (see, e.g., Figure~12
of Ha\c{s}egan et~al. 2005), so it is clear that the
bulk of the distribution lies in the regime accesible to us.

We have devised an analytical distribution function ${\cal J}$ with four
parameters which is able to describe the observed 
distribution of $\log \hat{r}_h$,
such  that the probability of finding a GC with $\log \hat{r}_h$ in an 
interval of width $d(\log \hat{r}_h)$ around $\hat{r}_h$ is given by
${\cal J}(\log \hat{r}_h)\,d(\log \hat{r}_h)$. The form of this 
function is
\begin{equation}
{\cal J}(\log \hat{r}_h|\Theta) = f {\cal G}(\log \hat{r}_h|\mu,\beta_1) + 
(1-f) {\cal G}(-\log \hat{r}_h|-\mu,\beta_2)
\label{eq:J}
\end{equation}
\noindent where

\begin{equation}
{\cal G}(x|\mu,\beta) = \frac{1}{\beta}\exp[-(x-\mu)/\beta -\exp(-(x-\mu)/\beta)],
\end{equation}

\noindent$\Theta \equiv (\mu,\beta_1,\beta_2,f)$, $\beta > 0$ and
$0 \le f \le 1$\footnote{The function ${\cal G}$ can be derived
as the distribution of extreme values for independent
identically distributed random
variables. While this fact might hold some clue as to the
physical basis for the observed distribution, we use it
here strictly as a statistical modelling
function without an {\em a priori} physical justification.}.

The equivalent form of this distribution in terms of 
$\hat{r}_h$, such that the probability
of finding a GC with $\hat{r}_h$ in an interval $d\hat{r}_h$ around $\hat{r}_h$ is
given by $\tilde{\cal J}(\hat{r}_h)\,d\hat{r}_h$, is
%(related by ${\cal J'}(x) = (\ln (10) x)^{-1}{\cal J}(x)$):

\begin{equation}
\tilde{\cal J}(\hat{r}_h|\tilde{\Theta}) = \tilde{f} \tilde{\cal G}(\hat{r}_h|\tilde{\mu},\tilde{\beta}_1) 
+ (1-\tilde{f}) \hat{r}_h^{-2} \tilde{\cal G}(\hat{r}_h^{-1}|\tilde{\mu}^{-1},\tilde{\beta}_2)
\label{eq:Jt}
\end{equation}

\noindent where

\begin{equation}
\tilde{\cal G}(x|\tilde{\mu},\tilde{\beta}) = 
\frac{1}{x\,\tilde{\beta}}\left[\frac{x}{\tilde{\mu}}\right]^{-1/\tilde{\beta}}
\exp (-(x/\tilde{\mu})^{-1/\tilde{\beta}}),
\end{equation}

\noindent $x,\tilde{\mu},\tilde{\beta} > 0$, $0 \le \tilde{f} \le 1$   
and the parameters $\tilde{\Theta} \equiv (\tilde{\mu},\tilde{\beta}_1,\tilde{\beta}_2,\tilde{f})$
are related to the ones
in ${\cal J}$ by the relations
$\tilde{\mu} = 10^\mu$, $\tilde{\beta}_{1,2} = \ln(10) \beta_{1,2}$ and
$\tilde{f}=f$. These functions
satisfy $\int_{-\infty}^{\infty} {\cal J}(\log r_h)\,d(\log r_h) =  \int_0^\infty \tilde{\cal J} (r_h)\,dr_h = 1$.

We have obtained the best-fit parameters $\mu,\beta_1,\beta_2$ and $f$
via maximum-likelihood for the various subsamples shown in 
Figure~\ref{fig:rhdist} as well as for the combined sample of GCs.
The resulting functions, convolved with a Gaussian with $\sigma=0.035$
in order to faithfully compare them with the density estimates,
are shown in Figure~\ref{fig:rhdist} with 
a thick dashed curve. The function is able to produce a close match
to the observed distributions. 

The maximum-likelihood parameters obtained for each galaxy subsample
are fairly consistent with the parameters inferred
for the combined sample of all galaxies. This is true despite the fact
that uncertainties arising from the effect of distance
uncertainties in combining the $\hat{r_h}$ values for several galaxies 
are not included in the formal error estimates. 
The parameters for the combined
sample, with uncertainties reflecting the dispersion
of the values inferred for all galaxy color bins, are
\begin{equation}
\begin{array}{rcc}
\mu & = & 0.407\pm0.013 \\
\beta_1 & = & 0.117\pm0.005 \\
\beta_2 & = & 0.078\pm0.011 \\
f & = & 0.7\pm0.1 \\
\end{array}
\end{equation}
and
\begin{equation}
\begin{array}{rcc}
\tilde{\mu} & = & 2.55\pm 0.08 \\
\tilde{\beta_1} & = & 0.27\pm 0.012 \\
\tilde{\beta_2} & = & 0.18\pm 0.025 \\
\tilde{f} & = & 0.7\pm0.1 \\
\end{array}
\end{equation}
We conclude that the observed distribution of $\hat{r}_h$ is well represented
by single distributions of the form 

\begin{equation}
{\cal J}(\log \hat{r}_h|\mu=0.407,\beta_1=0.117,\beta_2=0.078,f=0.7)
\label{eq:bf}
\end{equation}

\noindent or, equivalently,

\begin{equation}
\tilde{\cal J}(\hat{r}_h|\tilde{\mu}=2.55,\tilde{\beta}_1=0.27,\tilde{\beta}_2=0.18,\tilde{f}=0.7).
\end{equation}

\begin{figure}
\plotone{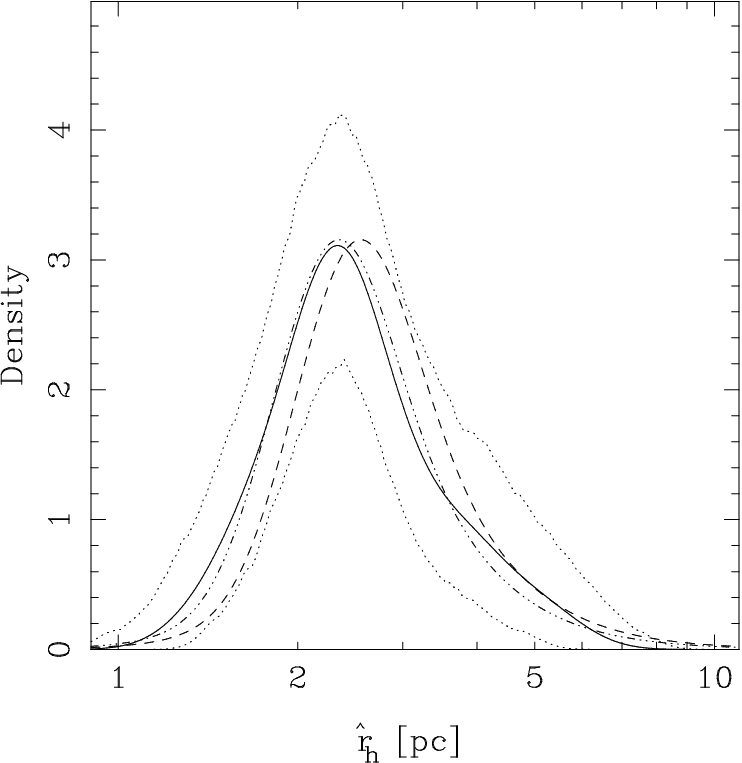}
\caption[]{ Distribution of $\log \hat{r}_h$ for Galactic GCs
satisfying $M_V \le -7.4$ mag and $0.75 < r_h < 10$ pc (66 GCs).
The solid curve shows a Gaussian kernel density estimate
with a bandwidth $h=0.06$, while the dotted curves shows its 99\%
confidence bands. The dashed line represents the distribution
given in equation~\ref{eq:bf} while the dash-dot-dot curve
represents the same curve with $\mu=0.37$.
}
\label{fig:mwrhdist}
\end{figure}

We now examine the applicability of the proposed function to the case
of the Galactic GC system.
In Figure~\ref{fig:mwrhdist} we show a Gaussian
kernel density estimate with $h=0.06$ of the distribution
of $\log{\hat{r}_h}$ for all GCs in our Galaxy satisfying
$\hat{r}_h < 10$ pc and $M_V < -7.4$ mag, along with the
99\% confidence bands. The individual $r_{h,i}$ 
have been corrected as described in \S~\ref{sec:distance}. We also
show the best-fit form of ${\cal J}$ shown in equation~\ref{eq:bf}
and convolved with a Gaussian with $\sigma=0.06$ as a dashed curve.
Without any adjustment the form of the $\hat{r}_h$ distribution
determined from our sample of early-type galaxies
is able to reproduce satisfactorily the form
observed in the Milky Way although the mean value
is differs slightly.\footnote{This mismatch would be reduced
if we had used the distance inferred to Virgo from the direct
geometric calibration using the Galactic GC system} The dash-dotted curve
shows the same curve after setting $\mu=0.37$, and shows
a very good agreement.
This exercise supports the flexibility of the proposed form for
describing the GC size distributions in galaxies of widely different
mass and morphological type. 

It is worth emphasizing that we have only modelled the {\it observed}
distribution of $\hat{r}_h$, which is the convolution of the 
intrinsic distribution with the error distribution. 
If the measurement uncertainties
are large compared to the width of the underlying 
distribution, then the observed distribution might be telling us 
more about the measurement errors than the actual
distribution of $\hat{r_h}$. However, as noted in \S~\ref{sec:obs},
the typical random uncertainties for our sample are of order 
$0\farcs003$ or $\approx 0.25$ pc at the mean distance of
Virgo, with the systematic uncertainties estimated
to be of the same order. The magnitude of the errors will
thus affect the width of the observed peak to some extent, 
although the extended, high-$\hat{r_h}$ tail will be largely
unaffected.

\begin{figure}
\plotone{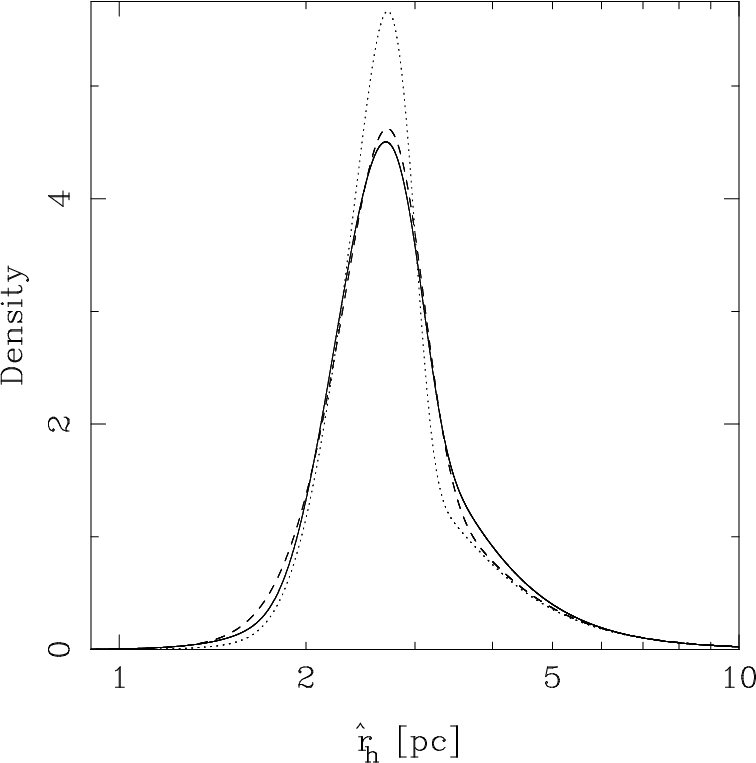}
\caption[]{ The solid line shows
the best-fit function ${\cal J}$ to the $r_h$
distribution of VCC1316 (M87 = NGC~4486). The dotted line shows
a function of the same form which best matches the observed
data for M87 after convolution with the average random
uncertainty in the $r_h$ measurement (assumed to be a Normal
distribution with $\sigma=0.25$ pc). The convolved
function is shown as the dashed line.
}
\label{fig:rhdistconv}
\end{figure}

To investigate the effect of random uncertainties
in a more quantitative way, we have taken the
best-fit function of the form of equation~\ref{eq:Jt}  
for VCC1316 (M87=NGC~4486) and determined the best match, via
least squares, of a function with this same form but convolved
with a Normal distribution with $\sigma=0.25$ pc in order 
to simulate the effect of the errors. The
use of a single galaxy for this comparison ensures that we
consider only random uncertainties: i.e., 
we are probing only the shape of the distribution,
and not the absolute value of its location parameter.
Figure~\ref{fig:rhdistconv} shows the result of this 
exercise. This figure demontrates
that the inferred intrinsic distribution shares the same features
as the one after convolution, suggesting that the observed distribution
is indeed revealing intrinsic properties. 
Additionally, the fact that ${\cal J}$ also provides a close match to the 
Galactic data, which has a completely independent error 
distribution, further shows that the form of ${\cal J}$ traces the
intrinsic properties of the $\hat{r_h}$ distribution.

Strictly speaking, the above analysis has relied on GCs that are brighter
than the peak of the GC luminosity function, so 
our results apply in that regime only. While
it is reasonable to expect the distribution to apply 
to the less luminous GCs, this claim needs to be tested.
Indeed, if we considered all Galactic GCs in
Figure~\ref{fig:mwrhdist}, then the agreement would be somewhat 
worsened due to the presence of a larger proportion of faint, extended
clusters. While the form of the distribution is still
well described by ${\cal J}$, the shape and location parameters need to be 
changed somewhat from those presented in equation~\ref{eq:bf}.
In particular, the extended high-$\hat{r_h}$-tail becomes more 
important when the fainter GCs are included in the analysis, which is reflected by the
larger $\beta_1$ (i.e., 0.16, as opposed to the value of 0.12
obtained when only the bright GCs are considered).

The form of the distribution presented above should serve
are a useful constraint for  models of GC formation and
numerical simulations of their dynamical evolution. 
Any viable picture of star formation in clusters should 
produce an observed size distribution that is consistent with the form
of ${\cal J}$, at least for GCs brighter than the luminosity
function turnover. 
The fact that its form seems to be uniform across the wide range
in luminosity and color spanned by ACSVCS program galaxies strongly
suggests that some quite generic mechanism determines ${\cal J}$, 
while the location of the distribution is set by
the local properties of the GC system and/or its host galaxy. 

\section{Summary \& Conclusions}

We have analysed the half-light radii of GCs in early-type galaxies 
using the exceptional dataset 
provided by the ACS Virgo Cluster Survey. We have presented
evidence that the average half-light radii $\langle r_h \rangle$ of GCs
belonging to early-type galaxies
increases with galactocentric distance or, alternatively,
with decreasing surface brightness.
This dependence is significantly
shallower than that observed in our Galaxy. For the
first time, we report an increase in $\langle r_h \rangle$
with decreasing galaxy color ($\sim$ decreasing luminosity), 
albeit with modest statistical significance. In agreement
with previous observations, we find strong evidence
that $\langle r_h \rangle$ is independent of the 
luminosity (mass) of GCs. We also find $\langle r_h \rangle$
to be $\sim 17 \%$ smaller for red (metal-rich) than for
their blue (metal-poor) counterparts. This
trend is better explained by a mechanism that is
intrinsic to GCs rather than to projection effects. As such,
the observations favor the picture in which this difference
follows from the combined effects of mass segregation
and the metallicity dependence of stellar lifetimes under
the assumption of invariant average half-{\it mass} radii 
(Jord\'an 2004).

We have discussed the predictions of two simple pictures
for the origin of the $r_h$ of GCs: one based in the idea
that they are determined by the average ambient pressure
surrounding proto-GC clouds, and other that $r_h$ is related
to the tidal radius set by the gravitational field of the
host galaxy. Both pictures are roughly consistent with the
observed behavior, $\langle r_h \rangle \sim L^0$,
although some elements of the theoretical framework are
lacking in both scenarios. While our level of understanding
of GC formation is still in its infancy, it is clear that
the observations of the $r_h$ of GCs can provide important
insights into the physical mechanisms which have shaped
their formation and evolution.

We have studied the form of the observed distribution of GC
half-light radii in early-type galaxies. Once corrected
for a variety of dependencies on color, radius and surface
brightness, the form of this distribution is remarkably
consistent across the range of color and luminosity spanned
by our sample of galaxies. We introduce simple analytic
expressions (equations~\ref{eq:J} and \ref{eq:Jt}) which 
satisfactorily describe the form of the $\hat{r}_h$ distribution
in our sample galaxies and in the Milky Way. Future theories
of GC formation and evolution should be able to produce
results consistent with this form when predicting the observed
distribution of $r_h$.

Once corrected for dependencies on color and
surface brightness, the mean half-light radii of GCs can be
used as a standard ruler for distance estimation. Specifially,
we find a constant value of
$\langle \hat{r_h} \rangle =2.7\pm0.3$ pc for a GC
with color $(g-z) = 1.2$, in a galaxy with
color $(g-z)_{\rm gal}=1.5$, and at an underlying $z$-band
surface brightness of $\mu_z = 21$ mag arcsec$^{-2}$.
A first attempt at a geometric calibration of this method based 
on the Galactic GC system gives an independent distance to the
Virgo Cluster of 16$\pm$2.3 Mpc.
While these conclusions rest heavily on two datapoints besides
the bulk of galaxies in Virgo (VCC~731 and NGC~4697), the ongoing ACS
Fornax Cluster Survey (Jord\'an et~al., in preparation), which is
similar to the ACSVCS but focusing on 43 early-type members of the
Fornax Cluster, will be invaluable in this regard by adding many
additional datapoints at $D\sim19$ Mpc (Tonry et~al. 2001).

While not as efficient some other distance indicators,
such as SBF, the use of $\langle \hat{r}_h \rangle$ 
as a distance indicator could, in principle, offer
some advantages that could render it a valuable complement
to other GC-based method such as the GC luminosity function.
For instance, half-light radii are expected to remain roughly constant during the
evolution of a GC. This is certainly not the case for luminosity, which
will be affected by age and metallicity, quantities
which are not necessarily known {\it a priori}. Additionally, it is 
not necessary to go past the turnover of the luminosity
function if enough clusters to accurately determine $\langle \hat{r_h} \rangle$
can be accumulated up to brighter magnitudes.
Assuming one could measure
$\langle \hat{r}_h \rangle$ up to $1/4$ of the PSF FWHM of 
a diffraction-limited telescope of diameter $d$ at a wavelength of $1\mu$,
this technique could be applied out to distances of $\sim 50$ Mpc ($d=6$m), 
$\sim 250$ Mpc ($d=30$m) and $\sim 850$ Mpc ($d=100$m), illustrating
the prospect for its implementation with future observing facilities. In
practice, the limitation will be most likely set by the difficulty of 
accurately characterizing the PSF over a sufficiently large field of view.

\acknowledgements

P.C. would like to thank the European Southern Observatory 
and especially Bruno Leibundgut for their hospitality during the preparation of this paper.
Support for programs GO-9401 and GO-10003 was provided through grants from
the Space Telescope Science Institute, which is operated by the Association
of Universities for Research in Astronomy, Inc., under NASA contract NAS5-26555.
P.C. acknowledges additional support provided by NASA LTSA grant NAG5-11714.
M.M. acknowledges additional financial support provided by the Sherman
M. Fairchild foundation. D.M. is supported by NSF grant AST-020631,
NASA grant NAG5-9046, and grant HST-AR-09519.01-A from STScI.
M.J.W. acknowledges support through NSF grant AST-0205960.
This research has made use of the NASA/IPAC Extragalactic Database (NED)
which is operated by the Jet Propulsion Laboratory, California Institute
of Technology.

\appendix

\section{KINGPHOT: An Algorithm for the Measurement of GC Photometric and
Structural Parameters Using HST/ACS Images}
\label{app:kingphot}

Measurements of the photometric and structural parameters of GCs in the
ACSVCS, including their half-light radii, were carried out with code
specifically written for these purposes. This code, hereafter referred to
as KINGPHOT, was designed to perform efficient and accurate
$r_h$ measurements for the thousands of GCs detected in the ACS Virgo 
Cluster Survey. The primary technical challenge in the design of the
code is to recover $r_h$ for the case of marginally resolved objects:
i.e., objects with intrinsic sizes comparable to, or smaller than,
the instrumental PSF. In this appendix, we outline the methodology
used by the code and present some simulations that illustrate
its performance.

What is observed at the detector is an integral over each pixel of the 
PSF-convolved object; note that we include in the term PSF the effects
of the telescope optics and detectors. The exact nature of the physical
effects that give rise to the observed light distribution are of no
concern to us here, as long as we have an empirical way of determining
the final distribution on the detector of the light from a celestial
point source. With this definition of the PSF, ${\cal P}$, 
the observed flux $O$ at a given pixel in the detector is given by

\begin{equation} 
O = \int\int_{\rm pixel} dx_1dx_2 [{\cal P} \otimes {\cal F}](x_1,x_2) 
\label{eq:conv}
\end{equation}

\noindent where ${\cal F}$ is the flux distribution of the object
and $\otimes$ denotes a convolution.

If ${\cal F}$ can be described 
parametrically by a set of parameters $\bf{\Theta}$, and if a model
for ${\cal P}$ is available, then the parameters in $\bf{\Theta}$ can
be simply estimated by $\chi^2$ minimization of the observed
two-dimensional light distribution minus a predicted model.
Because the family of steady-state dynamical models presented
by King (1966) are known to provide an excellent match to the observed
light profiles of most Galactic GCs, we have chosen
to parameterize ${\cal F}$ by these models. King  models are
a one-parameter family indexed by the central potential, $W_0$,
or equivalently, by the concentration, 
$c\equiv \log_{10} (r_t/r_c)$, where $r_t$ is the
tidal radius (the point at which the density is zero) and 
$r_c$ is the core radius (the point at which the density 
falls to roughly half of its central value). In addition to
$c$, there are two arbitrary scaling factors: the total magnitude, $m$, and
one related to spatial scale. The latter is typically taken to the core
radius, but we choose to work with the half-light
radius $r_h$. For King models, the ratio $r_h/r_c$ is uniquely determined by the 
concentration, so the choice of either is equivalent.

There is no analytic expression for the spatial density 
corresponding to the lowered Maxwellian phase space distribution
assumed by King (1966), but the
density is easily obtained by an integration of Poisson's 
equation. It would be prohibitively slow to compute this density
profile by solving Poisson's equation during the $\chi^2$ minimization,
as this would entail the computation of hundreds of 
models for each of the $\sim$ 10$^4$ GC candidates in the ACSVCS. 
Instead, a library of King-model density profiles was constructed, computed
in a grid of 300 equally spaced points in $W_0$ ranging from $W_0=1$ to 
$W_0=12$, which corresponds to the range $0.296 < c < 2.739$. This library 
is read into memory at the start of KINGPHOT and a model
of a given $c$ is obtained by spline interpolation between
the two closest models in the library. 

To represent ${\cal P}$, we adopt the method implemented in the DAOPHOT II
program (Stetson 1987; 1993).  We refer the reader to Stetson (1987) for 
the details of the PSF representation adopted. 
The PSFs for the current work were determined
empirically using moderately crowded fields in the Galactic
GC 47 Tucanae as described in Jord\'an et~al. (2004a). 

Given the adopted King model, the corresponding model parameters, and the
PSF model, the predicted light distribution at the detector can be 
obtained by calculating the convolution in equation~\ref{eq:conv}.
The convolution is performed via a fast Fourier transform
using routines from  the Fastest Fourier Transform in the West
library (FFTW; Frigo \& Johnson 2005)

Let us denote the observed data at a pixel $(i,j)$ by 
$d_{ij}$, and the prediction of a PSF-convolved King model
centered at position $(x_0,y_0)$ in the detector by 
$O_{ij}(c,m,r_h,x_0,y_0)$. The parameters $(c,m,r_h,x_0,y_0)$
are estimated by minimizing the value of $\chi^2$ given
by 

\begin{equation}
\chi^2 = \sum_{\{|(i,j)| < r_{\rm fit}\}} [w_{ij}(d_{ij} - O_{ij}(c,m,r_h,x_0,y_0))/\sigma_{ij}]^2
\end{equation}

\noindent where $\sigma_{ij}$ is the uncertainty of $d_{ij}$ 
and $w_{ij}$ are radial weights. 
The value of $\sigma_{ij}$ used by KINGPHOT is the expected 
$rms$ in the data image given by equation (2) of Jord\'an et~al. (2004a).
In the above summation, $\{|(i,j)| < r_{\rm fit}\}$ denotes all pixels
within a distance $r_{\rm fit}$ of $(x_0,y_0)$. The weights $w_{ij}$ 
are set to the value of the function $5 / (5 + 1/(r_{\rm fit}^2 /r_{ij}^2 -1))$, where
$r_{ij}$ is the distance of the pixel's center to $(x_0,y_0)$. These radial weights
are included to assist convergence as discussed in Stetson (1987). 
Initially, the parameters for all GCs were determined using
$r_{\rm fit}=4$ pixels. Our simulations show that for extended
clusters, with $r_h \gae r_{\rm rfit}/2$, the measured
$r_h$ starts to suffer from biases; for larger GCs,
we increased the value of $r_{\rm rfit}$ and redetermined the best-fit
parameters. This was repeated until the condition $r_h < r_{\rm rfit}/2$
is met or $r_{\rm fit}=15$ pixel. This condition is satisfied for
all GCs considered in this analysis.

Minimizations were carried with CERN's {\tt Minuit} package. The minimization
was done in two steps. First, a Simplex (Nelder \& Mead 1965) 
minimization algorithm was used, followed by a variable metric method with
inexact line search (MIGRAD). In the Simplex minimization the concentration
is held fixed at $c=1.5$, a typical value for Galactic GCs (Harris 1996).
The starting values for the parameters $(m,x_0,y_0)$ are obtained from SExtractor
(Bertin \& Arnouts 1996) while $r_h$ is set to $0\farcs0385$, roughly the angle
subtended by $3$ pc at the Virgo distance.  After the Simplex minimization, 
the center $(x_0,y_0)$ is fixed while $c$ is allowed to vary.
Before starting the MIGRAD minimization, a grid search is done in $(c,r_h)$ 
and the minimization is started at the parameters that give the lowest $\chi^2$ in
this grid search. This procedure is done to avoid biasing the final result by 
the choice of fixing $c$ in the Simplex minimization and to avoid
converging to a local, rather than a global, minimum. 
After MIGRAD is run, the best
fit parameters are recorded. To estimate the uncertainties a bootstrap 
procedure is used. From the oberved data $d_{ij}$ within the fitting radius,
a random set of $d_{ij}$ is generated and the
minimization is re-run, 50 times, with this new set. The output parameters
recorded, and the $rms$ of the recovered parameters is taken to represent
the uncertainties.

\begin{figure}
\epsscale{0.75}
\plotone{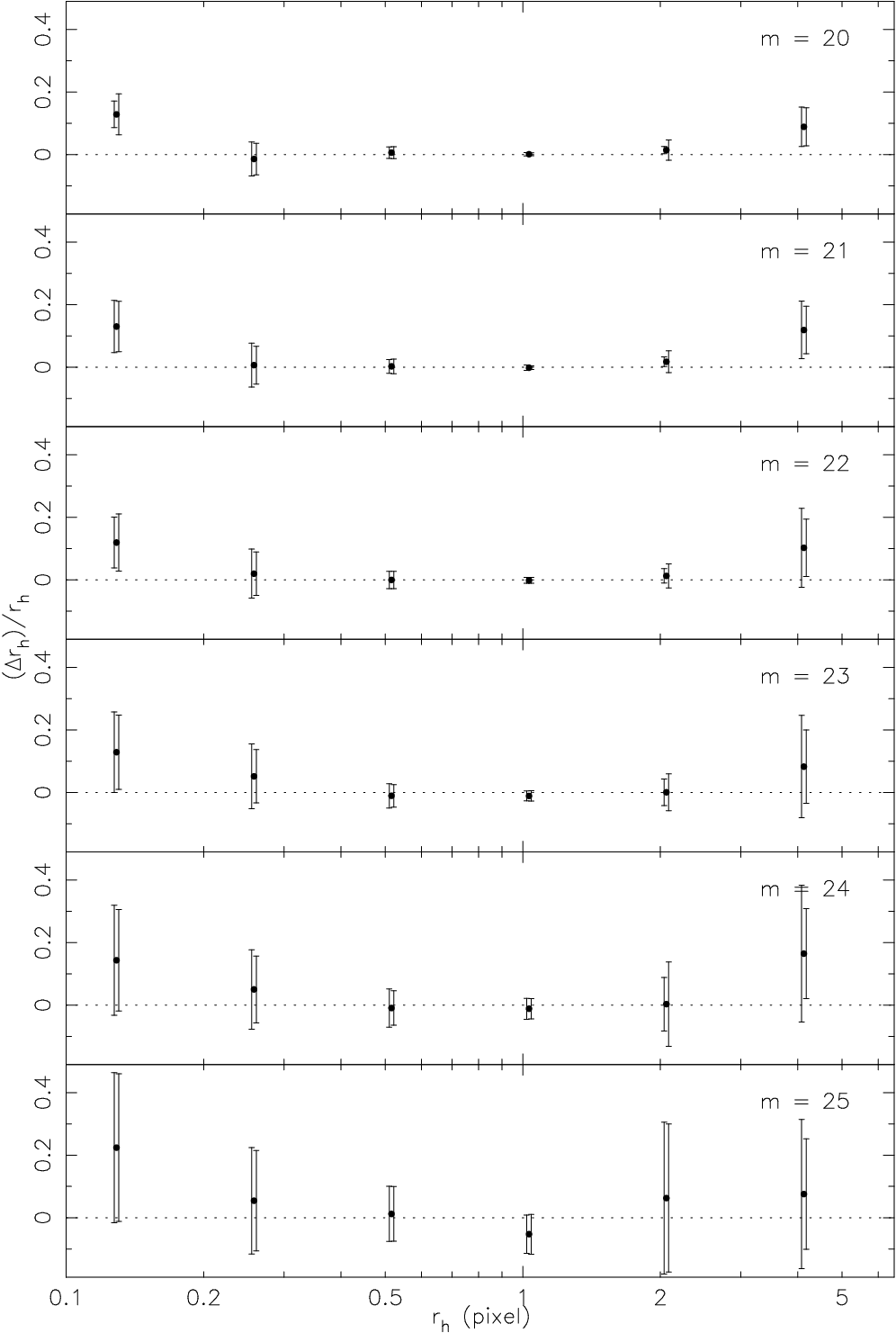}
\caption{$\Delta r_h / r_h$ versus $r_h$ for a set of KINGPHOT runs
on simulated GCs. The instrumental magnitude $m$ (for a 1s exposure)
of the simulated GCs
is indicated in each panel. For each simulated $r_h$, we plot the 
fractional mean difference between the measured and input $r_h$, $\Delta r_h / r_h$.
We also show two sets of error bars for each point: the left one is the 
observed rms of $\Delta r_h / r_h$ while the right one is the average
uncertainty computed by the code. The fitting radius was set to $r_{\rm fit}=4$ pixels. 
\label{fig:app}
}
\end{figure}

To test the performance of this code, it has been run on
a set of simulated GCs to which an appropriate
amount of instrumental and Poisson noise has been added.
For all combinations of $(c,r_h)$ in the grid $c=0.5,0.7,\ldots,2.1,2.3$ and
$r_h=0.5,1,2,4,8,16$ pc (at the distance of Virgo), 15 objects were 
simulated and the best-ft parameters recorded. This was done for
instrumental magnitudes of $m=20,21,22,23,24,25$ (for a 1s exposure),
with a fitting radius of $r_{\rm fit} = 4$ WFC pixels. The relevant results
of these simulations are summarized in Figure~\ref{fig:app}, where 
we show the average $\Delta r_h / r_h$ (for all simulated $c$ values). Here,
$\Delta r_h$ is the
difference between the recovered and input $r_h$,  
Two sets of error bars are shown for each point, slightly displaced.
The one to the left shows the observed $rms$ of $\Delta r_h / r_h$ while
the one to the right indicates the average uncertainty of
$\Delta r_h / r_h$ returned by KINGPHOT.
Instrumental magnitudes can be transformed into $z$-band ones by
$m_z \simeq m - 0.14$. Thus, only the four uppermost panels of
Figure~\ref{fig:app} are relevant to the present work (due to the
construction of the sample; see Equation~\ref{sel1}).

We conclude on the basis of these simulations that the performance of KINGPHOT
is, for observational material similar to the ACSVCS, largely unbiased for 
$r_h \gae 1$ pc. At the same time, it is apparent that, for $r_h \lae 1$ pc, 
there are always systematic biases which are of order $\lae 20\%$. It is
also apparent is that a bias arises when the intrinsic
$r_h$ approaches $r_{\rm fit}$, as is revealed by the rightmost point
in each panel. These bias, however, is avoided if $r_h \lae 0.5 r_{\rm fit}$,
so it is necessary that this condition is satisfied {\em post-facto} by iteratively
adjusting $r_{\rm fit}$ as we have done (see above). Finally, the uncertainties
returned by KINGPHOT agree very well with the observed dispersion on the 
$\Delta r_h / r_h$ values, and thus we conclude that our bootstrap procedure
estimates the uncertainties properly.
Needless to say, such simulations can provide information only 
on the intrinsic performance of the code, and neglect sources of error
arising from inaccuracies in the assumed 
forms for ${\cal F}$ or ${\cal P}$. 
Uncertainties arising from systematic errors of this sort can be
estimated from the independent measurements in the two bandpasses (see \S2).

\clearpage

\end{document}